\documentclass[11pt]{article}
\pdfoutput=1

\usepackage{bm,wrapfig,float,array,mathtools}
\usepackage{amsfonts}
\usepackage{amssymb}
\usepackage{cite}
\usepackage{amsmath}
\usepackage{color}
\usepackage{graphicx}
\usepackage{hyperref}
\usepackage{float}
\usepackage[T1]{fontenc}
\usepackage[utf8]{inputenc}
\usepackage{alphabeta}
\usepackage{fancyvrb}
\usepackage[dvipsnames]{xcolor}
\usepackage{bold-extra}
\usepackage{lmodern}
\usepackage{cleveref}
\usepackage[normalem]{ulem}
\usepackage{tikz-cd}
\usepackage{array}
\usepackage{tensor}
\usepackage{comment}
\usepackage{diagbox}

\usepackage{caption}
\graphicspath{ {figures/} }

\usepackage{tikz}
\tikzset{gauge/.style={circle, draw, line width=.2mm, inner sep=2pt,minimum size=3em}}
\tikzset{flavor/.style={rectangle, draw, line width=.2mm, inner sep=3pt}}
\tikzset{gline/.style={thick}}
\tikzset{doublearrow/.style={draw=black!75, color=black!75, double distance=5pt, thick}}

\newcommand{\bea}{\begin{eqnarray}}
\newcommand{\eea}{\end{eqnarray}}
\newcommand{\be}{\begin{equation}}
\newcommand{\ee}{\end{equation}}
\newcommand{\ba}{\begin{aligned}}
\newcommand{\ea}{\end{aligned}}
\newcommand{\bit}{\begin{itemize}}
\newcommand{\eit}{\end{itemize}}
\newcommand{\ben}{\begin{enumerate}}
\newcommand{\een}{\end{enumerate}}

\newcommand{\CFT}{\text{CFT}}
\newcommand{\AdS}{\text{AdS}}
\newcommand{\vol}{\text{vol}}

\newcommand{\lb}{\left(}
\newcommand{\rb}{\right)}
\newcommand{\lbb}{\left[}
\newcommand{\rbb}{\right]}

\newcommand{\tn}[1]{\textnormal{#1}}

\newcommand{\br}{\breve}

\newcommand{\half}{\frac{1}{2}}

\newcommand{\mb}{\mathbb}
\newcommand{\Z}{{\mathbb Z}}
\newcommand{\R}{{\mathbb R}}
\newcommand{\bC}{{\mathbb C}}

\renewcommand{\P}{{\mathbb P}}

\newcommand{\cA}{\mathcal{A}}

\newcommand{\cC}{\mathcal{C}}

\newcommand{\cG}{\mathcal{G}}

\newcommand{\cK}{\mathcal{K}}
\newcommand{\cL}{\mathcal{L}}
\newcommand{\cM}{\mathcal{M}}
\newcommand{\cN}{\mathcal{N}}

\newcommand{\cS}{\mathcal{S}}
\newcommand{\cT}{\mathcal{T}}

\newcommand{\cX}{\mathcal{X}}

\renewcommand{\d}{\text{d}}

\newcommand{\Stop}{S_\text{top}}

\def\ads#1{{\rm AdS}_{#1}}

\def\mc{\mathcal}

\def\wcX{\widetilde{\mathcal{X}}}

\begin{document}

\baselineskip=18pt  
\numberwithin{equation}{section}  
\allowdisplaybreaks  


\vspace*{0.8cm} 
\begin{center}
{{\huge Symmetry TFTs for 3d QFTs from M-theory}}

 \vspace*{1.5cm}
{\large Marieke van Beest$\,^{1}$, Dewi S.W. Gould$\,^2$, Sakura Sch\"afer-Nameki$\,^2$, Yi-Nan Wang$\,^{3,4}$}\\

 \vspace*{.2cm} 
\smallskip

{\it $^1$ Simons Center for Geometry and Physics, SUNY, \\
Stony Brook, NY 11794, USA}\\

\smallskip
{\it $^2$ Mathematical Institute, University of Oxford, \\
Andrew-Wiles Building,  Woodstock Road, Oxford, OX2 6GG, UK}\\

\smallskip
{\it $^3$ School of Physics,\\
Peking University, Beijing 100871, China\\ }

\smallskip

{\it  $^4$ Center for High Energy Physics, Peking University,\\
Beijing 100871, China}

\vspace*{2cm}
\end{center}

\noindent
We derive the Symmetry Topological Field Theories (SymTFTs) for 3d supersymmetric quantum field theories (QFTs) constructed in M-theory either via geometric engineering or holography. 
These 4d SymTFTs encode the symmetry structures of the 3d QFTs, for instance the generalized global symmetries and their 't Hooft anomalies. 
Using differential cohomology, we derive the SymTFT by reducing M-theory on a 7-manifold $Y_7$, which either is the link of a conical Calabi-Yau four-fold  or part of an $\text{AdS}_4\times Y_7$ holographic solution. 
In the holographic setting we first consider the 3d $\cN=6$ ABJ(M) theories and derive the BF-couplings, which allow the identification of the global form of the gauge group, as well as 1-form symmetry anomalies. 
Secondly, we compute the SymTFT for 3d $\mathcal{N}=2$ quiver gauge theories whose holographic duals are based on Sasaki-Einstein 7-manifolds of type  $Y_7 = Y^{p,k}(\mathbb{C}\mathbb{P}^2)$. The SymTFT encodes 0- and  1-form symmetries, as well as potential 't Hooft anomalies between these. Furthermore, by studying the gapped boundary conditions of the SymTFT we constrain the allowed choices for  $U(1)$ Chern-Simons terms in the dual field theory.

\newpage

\tableofcontents


\section{Introduction}
Symmetries and their anomalies have proven to be powerful tools in analysing quantum field theories (QFTs). 
Following the proposal in \cite{Gaiotto:2014kfa}, symmetries are now understood to be the  set of \textit{topological} operators in a given QFT. 
This generalization leads to a broadening of 
the paradigm of `symmetry', including higher-form symmetries \cite{Gaiotto:2014kfa}, higher-group symmetries \cite{Sharpe:2015mja,Tachikawa:2017gyf,Cordova:2018cvg,Benini:2018reh,Hidaka:2020izy,Hidaka:2021mml} and non-invertible symmetries -- the most recent progress being in higher-dimensional $d\geq 4$ theories \cite{Nguyen:2021naa,Heidenreich:2021xpr,Koide:2021zxj,Kaidi:2021xfk,Choi:2021kmx,Chatterjee:2022kxb,Roumpedakis:2022aik,Bhardwaj:2022yxj,Hayashi:2022fkw,Choi:2022zal,Kaidi:2022uux,Choi:2022jqy,Cordova:2022ieu,Bashmakov:2022jtl,Antinucci:2022eat,Choi:2022rfe,Damia:2022bcd, Bhardwaj:2022lsg, Lin:2022xod, Bartsch:2022mpm,  Apruzzi:2022rei, GarciaEtxebarria:2022vzq,Heckman:2022muc,Kaidi:2022cpf,Niro:2022ctq}. Since their recent inception, generalized symmetries in string theory and related theories have thus been studied extensively\footnote{For a sample list of references see
\cite{Tachikawa:2013hya, Morrison:2020ool,  GarciaEtxebarria:2019caf, Albertini:2020mdx, Bhardwaj:2020phs, Closset:2020scj, Closset:2020afy, Bhardwaj:2021pfz, Hosseini:2021ged,Apruzzi:2021vcu,Apruzzi:2021mlh,Cvetic:2021sxm,Buican:2021xhs,Bhardwaj:2021zrt,Braun:2021sex,Cvetic:2021maf,Closset:2021lhd,Gukov:2021swm,Closset:2021lwy,Yu:2021zmu,Sharpe:2021srf,Robbins:2021xce,Beratto:2021xmn,Bhardwaj:2021mzl,Tian:2021cif,DelZotto:2022joo,Bhardwaj:2022dyt,Hubner:2022kxr,Bhardwaj:2022scy,
DelZotto:2022fnw,Lee:2022spd,Carta:2022spy,DelZotto:2022ras,Argyres:2022kon,Heckman:2022suy,Cvetic:2022imb,Benedetti:2022zbb,Chatterjee:2022tyg,Lohitsiri:2022jyz,Pantev:2022kpl,Bolognesi:2022beq,Damia:2022rxw}. }.

The realization of QFTs within string theory has most of its utility when studying strongly coupled regimes of theories, either in geometric engineering or from a dual holographic perspective.
In some instances this provides otherwise inaccessible information about strongly coupled QFTs, and in particularly favorable circumstances even a framework for classification of particular types of (supersymmetric) QFTs.
The string theoretic realization has to capture some of the salient physical properties of the QFTs, in particular the generalized symmetries and their 't Hooft anomalies, which are robust under RG-flow.  
The symmetry structure of a QFT can be encoded in the so-called  \textit{Symmetry Topological Field Theory (SymTFT or Symmetry TFT)} \cite{Freed:2012bs, Gaiotto:2020iye, Apruzzi:2021nmk}, see \cite{Apruzzi:2022dlm, Apruzzi:2022rei, GarciaEtxebarria:2022vzq, saghar,Freed:2022qnc, Kaidi:2022cpf} for recent applications. 

In a nutshell, the SymTFT is a $(d+1)$-dimensional topological field theory, which upon reduction on an interval with topological boundary conditions on one side, and physical (non-topological) boundary conditions on the other, gives rise to the physical theory (and its anomaly theory). The SymTFT contains, for example, the BF-couplings 
 of the background fields for global symmetries and the couplings that give rise to 
 't Hooft anomalies.  We will shortly give a more thorough introduction in section \ref{sec:symtftintro}. 

The main observation in \cite{Apruzzi:2021nmk}, is that for QFTs that have a realization in string theory, the SymTFT can be derived from a supergravity approach. 
For a geometric engineering setup that corresponds to a dimensional reduction on a (non-compact) space $X$, the SymTFT is obtained by a suitable dimensional reduction on $\partial X$ -- and thus is naturally one dimension higher than the QFT that is being engineered. In order to capture subtle aspects such as finite group (higher-form) symmetries, the dimensional reduction is not a standard KK-reduction in supergravity, but we are required to utilize {\it differential cohomology} to capture background fields of finite group symmetries. Prior applications of differential cohomology to string/M-theory have appeared in \cite{Freed:2000ta, Hopkins:2002rd, Freed:2006mx, Bah:2020uev, Hsieh:2020jpj, Apruzzi:2021nmk}, and for a mathematical review see \cite{bar2014differential}. 

Closely related to this is that of holography, where the strongly-coupled regime of a superconformal field theory is realized in terms of string/M-theory on $\AdS_{d+1} \times X$ spacetime. In this case, the SymTFT can be interpreted as the topological couplings in the bulk supergravity on $\AdS_{d+1}$ (or in more general holographic setups). 
The most well-studied example of $\AdS_5\times S^5$ has the bulk coupling $N \int_{\AdS_5} B_2\wedge dC_2$, which is  precisely an example of such a BF-coupling for the 1-form symmetries of the dual 4d gauge theories (with gauge algebra $\mathfrak{su}(N)$) \cite{Witten:1998wy}. More precisely, 
the SymTFT in holography lives in the near-boundary region of the bulk and models the choice of global forms of the gauge group and the singleton sector \cite{Witten:1998wy, Belov:2004ht}. In terms of the formulation as generalized symmetries and SymTFTs, there has been much recent interest in the holographic literature  \cite{Witten:1998wy, Hofman:2017vwr,Iqbal:2020lrt, Bergman:2020ifi,Bah:2020uev, Bah:2021brs,Apruzzi:2021phx, Das:2022auy, Damia:2022bcd, Apruzzi:2022rei,Benini:2022hzx}, in particular for $\AdS_4/\CFT_3$ in  \cite{Bergman:2020ifi} for 3d $\mathcal{N}=6$ SCFTs of ABJM type \cite{Aharony:2008ug}.

The goal of this paper is to determine the SymTFT for 3d QFTs which either have a realization as geometric engineering in M-theory on an 8-manifold, or in terms of $\AdS_4/\CFT_3$ holographic setups in M-theory. 
These two constructions are closely related and we provide a systematic computational approach to determining the SymTFT in both cases. The main focus will be on conical 8-manifolds (with special holonomy) $\cX_8=\cC(Y_7)$ in setups with and without branes. 
Using differential cohomology in the supergravity reduction allows us to take into account the effects of torsion in the homology of $Y_7$, which is associated with a new set of background fields for finite higher-form symmetries.

For $Y_7$ a Sasakian 7-manifold we provide a prescription for computing the SymTFT coefficients explicitly, which correspond to secondary invariants in differential cohomology, from the intersection theory in the non-compact complex 4-fold $\cX_8$ \footnote{We assume that $\cX_8$ has a resolution, so that we can rely on a smooth model and intersection theory therein.}. We give detailed examples when the cone $\cX_8$ is toric, in particular for $\cX_8=\bC^4/\Z_k$ and $\cX_8=\cC(Y^{p,k}(\bC \P^2))$, where combinatorial formulas for intersection numbers can be explicitly computed. As such, we explain how physical anomaly coefficients and BF-terms are encoded in the geometric information of the toric diagram.
In summary, we will derive the SymTFT and give a procedure for computing the coefficients for 
\begin{enumerate}
\item Geometric engineering: M-theory on a singular, non-compact Calabi-Yau 4-fold $\cX_8=\cC(Y_7)$, i.e. $Y_7$ is a Sasaki-Einstein 7-manifold. 
\item Holography: $\AdS_4\times Y_7$ solutions of M-theory, which are dual to M2-branes probing $\cX_8=\cC(Y_7)$, where $Y_7$ is a Sasakian 7-manifold (Sasaki-Einstein when $\cX_8$ is a Calabi-Yau 4-fold). 
\end{enumerate}
For concrete applications, we will mostly focus on the holographic setups, leaving the exploration of geometrically engineered 3d QFTs for future work.
We first compute the SymTFT in the M-theory models dual to ABJM and ABJ theories. This relatively simple holographic setup is well-suited to demonstrate these new refined geometric methods while, at the same time, allowing for a match with known results from type IIA \cite{Bergman:2020ifi} in the case where discrete background torsional flux is turned off.
Finally, we apply this machinery in a much more subtle (and not completely fixed) duality of 3d $\mathcal{N}=2$ theories realized on M2-branes probing $\cC(Y^{p,k}(\mathbb{C}\mathbb{P}^2))$ \cite{Gauntlett:2004hh,Martelli:2008rt,Benini:2011cma}. By computing the SymTFT from the geometry, we obtain previously unknown anomalies for these theories. Furthermore, we will see that analysing consistent gapped boundaries of the SymTFT provides some further checks and balances to the proposed dictionary, coming from the spectrum of extended operators.

Generalized symmetries and their 't Hooft anomalies have a rich structure that has been studied field-theoretically from various angles in  e.g. in \cite{Cordova:2017vab, Hsin:2018vcg,  Eckhard:2019jgg, Hsin:2020nts, Bhardwaj:2022dyt}. Some of these results will be used later on to cross-check against our string theoretic results.

Let us summarize some of the main results.

\paragraph{ABJ(M). }
The  $\cN=6$ ABJM theories were conjectured in \cite{Aharony:2008ug} as a class of $U(N)_k \times U(N)_{-k}$ Chern-Simons matter theories with bifundamental matter, realized on $N$ M2-branes probing $\cC(S^7/\Z_k) = \mathbb{C}^4/\Z_k$. The addition of $b$ fractional M2-branes takes us to the ABJ variant \cite{Aharony:2008gk} with $U(N+b)_k \times U(N)_{-k}$ gauge group. The theories are conjectured to be holographically dual to M-theory on AdS$_4 \times S^7/\Z_k$ with $N$ units of 4-form flux over the external space. It was argued in \cite{Aharony:2008gk} that the presence of fractional branes gives rise to an additional $b$ units of torsional $G_4$ flux in the near-horizon limit. 

In \cite{Bergman:2020ifi}, generalized symmetry methods were used to derive a suite of gauge theories within the framework without fractional branes by considering different boundary conditions of a topological field theory in one dimension higher. In particular, these gauge theories have the same Lie algebra as the $U(N)\times U(N)$ ABJM theory, but different global forms of the gauge group.

In this work we use differential cohomology tools in M-theory to determine the BF-term
\be 
\frac{S_\tn{kin}}{2\pi}= \int_{\ads{4}} k B_2 \wedge dB_1+N B_2 \wedge F\,,
\ee 
which matches that found from IIA in \cite{Bergman:2020ifi}. We emphasize that whilst the BF-term is familiar, we must employ recent technology \cite{Apruzzi:2021nmk} to understand the geometric origin of the 1-form symmetry background. In this work, we make this link precise by reducing 11-dimensional supergravity on the \textit{torsional} components of the M-theory geometry. We identify the 1-form symmetry background as the reduction of $\br G_4$, a differential cohomology refinement of $G_4$, on the generator of $H^2(S^7/\Z_k,\Z)=\Z_k$.
Furthermore, we give a geometric derivation of a 1-form symmetry anomaly of the ABJ theories
\be 
-\frac{b}{2k} \int_{\ads{4}} B_2 \smile B_2\,.
\ee 
The anomaly is known from field theory \cite{Tachikawa:2019dvq}, but, to our knowledge, has not previously been detected from geometry.

\paragraph{$\cN=2$ Quiver Gauge Theories. } We also investigate a set of holographic quiver gauge theories on $N$ M2-branes probing a Calabi-Yau 4-fold $\cC(Y_7)$ which preserve $\cN=2$ supersymmetry. We take the 7-manifold to be $Y_7= Y^{p,k}(\mathbb{C}\mathbb{P}^2)$ \cite{Gauntlett:2004hh,Martelli:2008rt}, which are $S^3/\Z_p$ bundles over $\mathbb{C}\mathbb{P}^2$. In \cite{Benini:2011cma} the field theory duals to these geometries are proposed, where the authors furthermore consider the effect of wrapped M5-branes on non-trivial cycles in $H_3(Y^{p,k},\Z)$, corresponding to torsional $G_4$ flux. They derive a triangular Chern-Simons quiver with gauge group 
\be
\cG_{Y^{p,k}} = \prod_{i=1}^3 U(N_i)_{k_i} \,.
\ee
We compute all SymTFT terms arising from the differential cohomology reduction of 11-dimensional supergravity. We derive the BF-term using a combination of M-theory techniques, including the differential cohomology reduction, supplemented with a type IIA computation. Given the SymTFT, we can then realise all different global forms of the gauge group of the quiver by imposing boundary conditions on the gauge fields consistent with the SymTFT. We also derive a new 1-form symmetry anomaly for a subclass of these theories, which restricts certain choices of global form. Furthermore, we show that picking certain boundary conditions can induce the presence of residual mixed 0-/1-form symmetry anomalies in the boundary field theory. 

Finally, the field theories conjectured in \cite{Benini:2011cma} suffer from a parity anomaly. This gauge anomaly can be cured by turning on additional $U(1)$ CS-terms, which however are not completely determined by anomaly cancellation. Since the monopole operators and therefore the screening of line operators are sensitive to these additional CS-levels, the SymTFT can be used to constrain the additional terms in the Lagrangian.

The structure of this paper is as follows:
In section \ref{sec:symtftintro} we provide a brief  review of the concept of the SymTFT.
In section \ref{sec:SymTFT} we provide some background on differential cohomology and
compute a general expression for the SymTFT for  3d QFTs which can be constructed from M-theory on $\cX_8=\cC(Y_7)$ with and without branes. 
We then explain how to compute the coefficients in the SymTFT in section \ref{sec:symtftforSE7manifolds}, in particular in the case of toric $\cX_8$.
In section \ref{sec:ABJM} we apply the above technology to our first example: the 3d $\cN=6$ $((U(N+b)_k \times U(N)_{-k})/\Z_{m}$ ABJ(M) theories \cite{Aharony:2008ug,Aharony:2008gk,Tachikawa:2019dvq}. 
We next apply our technology to the $Y^{p,k}(\mathbb{C}\mathbb{P}^2)$ 3d $\cN=2$ quiver gauge theories of \cite{Gauntlett:2004hh,Martelli:2008rt,Benini:2011cma} in section \ref{sec:symtftholopart2}. 
In section \ref{sec:fieldtheorymatching} we discuss matching with field theory results of \cite{Benini:2011cma}. 
Finally, in section \ref{sec:outlook} we highlight various possible future directions.
We also provide some appendices. In appendix \ref{sec:isometries} we consider contributions to the SymTFT from background fields obtained by gauging spacetime isometries. In appendix \ref{sec:IIAanalysis} we use type IIA to conjecture the existence of an additional BF-term in the $Y^{p,k}$ case.

\section{The Symmetry Topological Field Theory}
\label{sec:symtftintro}

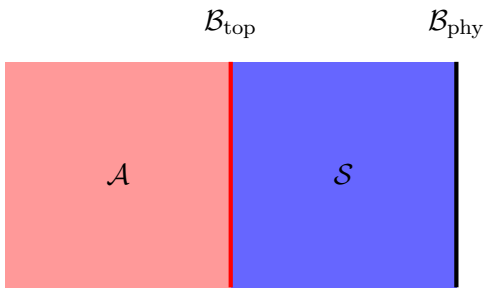
\begin{figure}
\centering
\begin{tikzpicture}
        \fill [red!40] (-3,0) rectangle (0,3);
        \fill [blue!60] (0,0) rectangle (3,3);
        \draw[red, ultra thick] (0,0) -- (0,3);
        \node[] (n3) at (1.5,1.5) {$\cS$};
        \draw[black, ultra thick] (3,0) -- (3,3);
        \node[] (n1) at (-1.5,1.5) {$\cA$};
        \node[] (n1) at (3,3.5) {$\mathcal{B}_{\text{phy}}$};
        \node[] (n1) at (0,3.5) {$\mathcal{B}_{\text{top}}$};
\end{tikzpicture}
\caption{
    The blue slab ("sandwich") is the SymTFT $\mathcal{S}$, with boundary conditions on the right ($\mathcal{B}_{\text{phy}}$) which are non-topological and to the left, which are topological $\mathcal{B}_{\text{top}}$. The red line is a gapped interface separating the symmetry theory from the anomaly theory $\cA$. The picture introduced earlier of $\cT$ being defined as a theory relative to $\cA$ is obtained by collapsing the blue sandwich. The anomaly theory describes anomalies for a particular choice of global structure of the QFT.
    }
\label{fig:anomalytheorydiagram}
\end{figure}

Central to our analysis is the \textit{Symmetry Topological Field Theory} (SymTFT). This will be, for the purpose of this paper, a $(d+1)$-dimensional topological field theory, which encodes salient symmetry structures of a $d$-dimensional QFT, obtained by an interval reduction. In particular it captures:
\begin{itemize}
    \item The choice of global structure of the gauge group.
    \item The 't Hooft anomalies of the higher-form and higher-group symmetries.
\end{itemize}
More generally the SymTFT will have topological boundary conditions (to the anomaly theory) and non-topological boundary conditions, which upon reduction along an interval result in the $d$-dimensional QFT. The SymTFT is an extension of the usual \textit{anomaly theory} picture in the following way. A $d$-dimensional QFT $\cT$ is relative to an \textit{invertible} theory which we call the anomaly theory. The gauge variation of the anomaly theory placed on a manifold with boundary will, by definition, exactly cancel the anomalous variation of the partition function of $\cT$. An invertible theory assigns a 1-dimensional Hilbert space to closed codimension-1 sub-manifolds in spacetime. This is the partition function of the QFT $\cT$, evaluated on the codimension-1 manifold.
Relaxing the invertible condition gives what is called the Symmetry Topological Field Theory. This allows for the assignment of a larger-than-one dimensional Hilbert space. Now the QFT has a vector of partition functions. This set of distinct partition functions encodes the possible choices of global structures of $\cT$. In particular, the QFTs encoded in these choices will have identical local physics but a distinct spectrum of extended operators. The setup is summarized in figure \ref{fig:anomalytheorydiagram}.

The concept of a SymTFT is in principle completely general and can be applied to capture the global structures of any given QFT\footnote{In a more categorical setting the SymTFT is the Drinfeld center of the symmetry category of the QFT.}. 
However, it is a particularly useful notion in the context of string theory since recent progress has showed that the SymTFT can be computed independently using geometric methods. 

\paragraph{Brane Constructions and Inflow. } Brane constructions in string theory provide a large class of examples of anomaly theories. Ambient space gauge anomalies are cancelled by worldvolume 't Hooft anomalies via so-called `anomaly inflow'. In particular, cutting out a neighbourhood around the branes, which act as sources of flux in the ambient string theory background, induces a boundary in the 10/11d geometry, rendering the full effective action no longer gauge invariant. In \cite{Bah:2019rgq,Bah:2020uev} it was explained that these anomalies, described by a $(d+1)$-dimensional TFT or $(d+2)$-dimensional anomaly polynomial, can be obtained by dimensional reduction of the topological terms of the 10/11d effective action.

\paragraph{Geometric Engineering. } In string theory constructions without branes, the notion of inflow becomes less clear. However it was argued in \cite{Apruzzi:2021nmk} that for compactifications on a $(D-d)$-dimensional cone $\cC(Y_{D-d-1})$ (with $D=10, 11$), dimensional reduction on the link space $Y_{D-d-1}$ remains a powerful tool in determing 't Hooft anomalies. The cases considered in \cite{Apruzzi:2021nmk} are 7d Yang-Mills and 5d SCFTs obtained from M-theory on singular Calabi-Yau spaces. The SymTFT is derived in both cases from dimensional reduction of the topological terms in the 11d supergravity action and is tested with non-trivial checks with known field theory computations for certain anomalies.

\paragraph{Background Fields from Cohomology. } 
The SymTFT in the cases discussed here is formulated in terms of background fields for various generalized symmetries. These correspond to massless gauge fields in the supergravity reduction, of which there are two sources: the reduction of the supergravity gauge potentials $C_n$ on the cohomology of the internal space $Y_{D-d-1}$, and the gauging of isometries of the geometry. First, considering \textit{continuous} symmetries (and specializing to $D=11$): expanding the M-theory $C_3$ field on representatives of the free part of the cohomology $H^p_{\tn{Free}}(Y_{10-d};\Z)$ gives rise to massless $(3-p)$-form gauge fields. Schematically, we write
\be
G_4 = dc_3+ \sum_i dc_2^i \wedge \omega_1^i + \sum_j dc_1^j \wedge \omega_2^j + \sum_k dc_0^k \wedge \omega_3^k\,,
\ee
where subscripts denote form degrees and the forms $\omega_p$ are representatives of the free parts of the $p^{\text{th}}$ cohomology group. Superscripts represent various components of the integral cohomology groups $H^p_{\tn{Free}}(Y_{10-d};\Z)$. The massless $q$-form gauge fields $c_q$ furnish background fields for \textit{continuous} $(q-1)$-form symmetries when fixed on the boundary.

\paragraph{Torsional Contributions.} An obvious extension is to consider \textit{finite} higher-form symmetries that arise from \textit{torsional} contributions to the cohomology of $Y_{10-d}$. Manifesting the associated discrete background gauge fields requires a reduction of $C_3$ on torsional cocycles: a problem beyond the scope of ordinary differential forms \footnote{Some attempts towards using standard harmonic forms were made in \cite{Camara:2011jg, Berasaluce-Gonzalez:2012abm}.}. This is where the framework of \textit{differential cohomology} $\br H (Y_{10-d})$ can be used to incorporate more general symmetry structures \cite{Apruzzi:2021nmk}. We include torsional contributions by lifting $G_4$ to differential cohomology and expanding as follows
\be
\br G_4 = \sum_{\alpha} \br B_3^{\alpha} \star \br t_1^{\alpha}+\sum_{\beta} \br B_2^{\beta} \star \br t_2^{\beta}+\sum_{\gamma} \br B_1^{\gamma} \star \br t_3^{\gamma} + \sum_\delta \br b^{\delta} \star \br t_4^{\delta} \,.
\ee
Here, $\br t_p^\alpha$ are differential cohomology lifts of generators of Tor$H^p(Y_{10-d};\Z)$ of torsional degree $\ell_p^\alpha \in \mathbb{N}$. We leave a detailed explanation of this notation and technology for later sections. Here, we wish only to demonstrate that the notion of `expanding $G_4$ in cohomology' is maintained. The fields $\br B_q^\alpha$ represent background fields for $\Z_{\ell_p^\alpha}$ higher-form symmetries. 
Crucially, including gauge fields of this new type allows for a whole new class of SymTFT couplings upon dimensional reduction. It is terms of this type in particular that we explore in this paper.

\paragraph{Background Fields from Isometries. }  \textit{Isometries} of the internal space $Y_{10-d}$ are another source of background fields.  Turning on 1-form gauge fields for isometries is equivalent to fibering the manifold $Y_{10-d}$  over the external space. At the level of the SymTFT, this procedure will generate new couplings involving these 1-form gauge fields. We consider a subset of such terms by gauging spacetime isometries within the \textit{free} subsector of the cohomology reduction, see appendix \ref{sec:isometries}. We leave the torsional components of this story to future work.

\paragraph{Holographic Theories.} 
Much progress has been made in recent years in identifying generalized symmetry structures in holographic correspondences. In this paper, we will discuss it in detail for $\AdS_4$ solutions. For now, let us simply exemplify the most basic aspect of this: 
 the choice of global structure for the boundary theory is encoded in a BF-term for the bulk gauge fields, which constrains the possible boundary conditions one can impose on the fields \cite{Witten:1998wy, Maldacena:2001ss,Belov:2004ht}. For example, in \cite{Apruzzi:2021phx, Apruzzi:2022rei} the following 5d bulk supergravity term was derived in the Klebanov-Strassler holographic solution \cite{Klebanov:2000hb}
\be
\label{eq:KS-BF}
\frac{S_{\text{BF}}}{2\pi} = \text{gcd}(N,M) \int_{\cM_5} b_2 \wedge d\cC_2 \,,
\ee
for integers $N,M$ and 2-form fields $b_2,\cC_2$. The equations of motion force $b_2,\cC_2$ to be flat gauge fields. The topological bulk operators $e^{i \oint b_2}$, $e^{i\oint \cC_2}$ are mutually non-local due to the BF-action since the two composite fields are canonically conjugate \cite{Witten:1998wy}. Now suppose the following boundary conditions are chosen:
\be
b_2 \text{ Dirichlet} \,, \quad \cC_2 \text{ Neumann} \,.
\ee
Then $e^{i \oint \cC_2}$ are the topological codimension-$2$ operators in 4d generating a $\Z_{\text{gcd}(N,M)}$ 1-form symmetry with charged lines given by the operators $e^{i \oint b_2}$ restricted to the boundary. Alternate choices of boundary conditions correspond to different boundary global symmetries or, equivalently, different choices of global form of the boundary field theory gauge group.
This type of analysis has been extend to a 3d example in \cite{Bergman:2020ifi} and 4d in \cite{Apruzzi:2021phx}. Anomalies in holographic theories have also been studied for example in \cite{Bah:2019rgq,Bah:2020uev}.

\section{SymTFT from M-Theory on $Y_7$}
\label{sec:SymTFT}
We derive the SymTFT of any 3d QFT that arises in M-theory, either as compactification on $\R^{1,2} \times \cC(Y_7)$, or holographically dual to $\ads{4} \times Y_7$. This is achieved by reducing the topological terms of 11d supergravity on both the free and torsional parts of the cohomology of $Y_7$. A caveat in this analysis is that the symmetries we will capture from this approach need to be manifest within the geometric realization. We focus our main attention on the dimensional reduction using differential cohomology. A generalization to equivariant differential cohomology is discussed in appendix \ref{sec:isometries}, where gauging of isometries is included, which can give rise to additional global symmetries of the QFTs.

\subsection{Reduction using the Free Part of Cohomology}
\label{sec:free}

Let us start by performing the reduction of M-theory on $\cM_{11}=\cM_4 \times Y_7$, using only the free part of the cohomology $H^p_{\tn{Free}}(Y_7;\Z)$ which gives rise to continuous gauge fields in the effective 4d theory. As discussed above, these massless modes are obtained by a Kaluza-Klein expansion of the 4-form flux $G_4$ on representatives of the cohomology of the internal space with integral periods. 
Their topological couplings arise from the 11d supergravity term
\be
\label{eq:CSaction} 
\frac{S_\text{11d}}{2\pi}= \int_{\cM_{11}} \lbb - \frac{1}{6} C_3 \wedge G_4 \wedge G_4-C_3 \wedge  X_8 \rbb \,.
\ee 
The 8-form characteristic class $X_8$ is constructed from the Pontryagin classes of the tangent bundle
\be
X_8 = \frac{1}{192} \left( p_1(T\cM_{11}) \wedge p_1(T\cM_{11}) - 4p_2(T\cM_{11}) \right) \,.
\ee
To derive the 4d topological couplings we consider the gauge invariant 5-form $I_5$, on an auxiliary 5d space, which is the derivative of the 4d topological Lagrangian
\be 
I_5=dI_4\,, \quad S_{4d}=2\pi \int_{\cM_4} I_4 \,.
\ee 
We identify $I_5$ as
\be 
\label{eq:I12}
I_5 = \int_{Y_7} I_{12}= \int_{Y_7} \left(-\frac{1}{6} G_4 \wedge G_4 \wedge G_4-G_4 \wedge X_8 \right)\,.
\ee 
Assuming $Y_7$ is connected, the betti numbers $b^r(Y_7)=\text{dim}H^r(Y_7,\R)$ satisfy $b^i(Y_7) = b^{7-i}(Y_7)$. We denote the associated closed $p$-forms by
\be 
\omega_p^i\,, \qquad p=0,\dots,7\,, \qquad i=0,\dots,b^p(Y_7)\,,
\ee 
with $\omega_0\equiv 1$. We expand the 4-form flux using these forms
\be
\label{eq:fluxexp} 
G_4= \sum_{p=0}^4 \sum_{i=0}^{b^p(Y_7)} g_{4-p}^i \wedge \omega_{p}^i\,.
\ee 
When considering particular solutions it will be convenient to have separated the background $G_4^\text{bg}$ supporting the vacuum from the dynamical fluctuations $G_4'$ around the solution:
\be 
G_4=G_4'+G_4^\text{bg}\,.
\ee
Imposing the Bianchi identity, we find that $g^i_q$ can locally be written as
\be 
\label{eq:fluxcomponents}
g_0^i\equiv\cN^i\,, \qquad g^i_q=dc^i_{q-1}\,, \quad q=1,2,3\,, \qquad g_4=dc_3+\cL \vol_{\cM_4}\,,
\ee 
with the background parametrised by
\be
\label{eq:fluxquant} 
\int_{\cC^i} G_4^\text{bg} =\cN^i \in \Z\,, \qquad  
\int_{Y_7} \star G_4^\text{bg} = \cL \in \Z\,, 
\ee 
where $\cC^i$ is a basis of 4-cycles in $Y_7$. We can therefore write the fluctuations
\be 
G_4'=  \sum_{p=0}^3 \sum_{i=0}^{b^p(Y_7)} dc_{3-p}^i \wedge \omega_{p}^i\,,
\ee 
and background
\be 
\label{eq:G4bg}
G_4^\text{bg}=\cL \vol_{\cM_4}+\sum_{i=0}^{b^4(Y_7)} \cN^i \omega_4^i\,.
\ee 
In the reduction of the CS-term $G_4^3$, the background flux over the external space will contribute metric-dependent terms (which belong to the scalar potential) that we neglect. 
Performing the reduction we find
\be 
\ba 
\int_{Y_7} -\frac{1}{6}G_4^3=&\sum_{ijk}\lb -\half K^{ijk} dc_2^i \wedge dc_0^j \wedge dc_0^k +\half \mathcal{K}^{ijk}  dc_1^i \wedge dc_1^j \wedge dc_0^k  + \mathbb{K}^{ijk} \cN^i dc_1^j \wedge dc_2^k \rb \\
+ &\sum_{ij} \mathfrak{K}^{ij} \cN^i dc_0^j \wedge dc_3\,,
\ea 
\ee 
where the intersection numbers are given by
\be 
\ba 
\label{eq:int}
K^{ijk}&= \int_{Y_7} \omega_1^i \wedge \omega_3^j \wedge \omega_3^k\,, \qquad &&\mathcal{K}^{ijk} = \int_{Y_7} \omega_2^i \wedge \omega_2^j \wedge \omega_3^k\,, \\
\mathfrak{K}^{ij}&=\int_{Y_7}  \omega_4^i \wedge \omega_3^j\,, \qquad &&\mathbb{K}^{ijk} = \int_{Y_7}  \omega_4^i \wedge \omega_2^j \wedge \omega_1^k\,.
\ea 
\ee 
To factorise the characteristic class $X_8$, we can employ the Whitney sum formula for Pontryagin classes defined on a product manifold \cite{10.2307/2044298}. Assuming the external space $\cM_4$ is orientable and spin, we obtain
\be\ba
p_1(T\cM_{11}) &= p_1(T\cM_4) + p_1(TY_7) \,,\\
p_2(T\cM_{11}) &= p_2(T\cM_4) + p_2(TY_7)+p_1(T\cM_4) \smile p_1(TY_7) \,.
\ea\ee
The second Pontryagin classes vanish on dimensional grounds. We can therefore write the 8-form characteristic class
\be
\label{eq:X8}
X_8 = - \frac{1}{96} p_1(T\cM_4) \smile p_1(TY_7)\,.
\ee
Together with the expansion \eqref{eq:fluxexp} we find
\be 
-\int_{Y_7} G_4\wedge X_8=-\frac{1}{96} \sum_{i=1}^{b^3(Y_7)} \lbb \int_{Y_7} \omega_3^i \wedge \ p_1(TY_7) \rbb dc_0^i \wedge p_1(T \cM_4)\,.
\ee 
Defining
\be 
C^i =\frac{1}{96} \int_{Y_7} \omega_3^i \wedge \ p_1(TY_7) \,,
\ee 
the gauge invariant 5-form is
\be 
\ba 
\label{eq:I5Free}
I_5= &\sum_{ijk}\lb -\half K^{ijk} dc_2^i \wedge dc_0^j \wedge dc_0^k +\half \mathcal{K}^{ijk}  dc_1^i \wedge dc_1^j \wedge dc_0^k  + \mathbb{K}^{ijk} \cN^i dc_1^j \wedge dc_2^k \rb \\
+ &\sum_{ij} \mathfrak{K}^{ij} \cN^i dc_0^j \wedge dc_3 - \sum_i C^i dc_0^i \wedge p_1(T \cM_4)\,.
\ea 
\ee 
Acting with an anti-derivative, we find
\be 
\ba 
\label{eq:I4Free}
I_4= &\sum_{ijk}\lb \half K^{ijk} dc_2^i \wedge c_0^j \wedge dc_0^k +\half \mathcal{K}^{ijk}  dc_1^i \wedge c_1^j \wedge dc_0^k  + \mathbb{K}^{ijk} \cN^i c_1^j \wedge dc_2^k \rb \\
- &\sum_{ij} \mathfrak{K}^{ij} \cN^i dc_0^j \wedge c_3 - \sum_i C^i c_0^i p_1(T \cM_4)\,.
\ea 
\ee 
Notice in particular the single derivative terms
\be 
I_4 \supset \sum_{ijk} \mathbb{K}^{ijk} \cN^i c_1^j \wedge dc_2^k - \sum_{ij} \mathfrak{K}^{ij} \cN^i dc_0^j \wedge c_3 \,.
\ee 
Such BF-terms constrain the possible boundary conditions that can be imposed on the pairs $(c_1^j,c_2^k)$ and $(c_0^j,c_3)$, which in turn dictates the global symmetries of the resulting field theory. For example, if $\mathbb{K}^{xyz} \cN^x \equiv m \neq 0$, giving $c_1^y$ Neumann (free) boundary conditions implies that $c_2^z$ must be fixed to a background value in $\Z_m$ in the boundary theory, giving rise to a $\Z_m^{(1)}$ global 1-form symmetry in the 3d theory. Exchanging the boundary conditions corresponds to gauging the full $\Z_m$ 1-form symmetry, and we obtain a $\Z_m$ 0-form symmetry instead.

\subsection{Review of Differential Cohomology}
\label{sec:diffcohoreview}
In \cite{Apruzzi:2021nmk} it was shown, by employing a description in terms of {\it differential cohomology}, that torsion in $H^p(Y_7;\Z)$ may give rise to additional couplings in the SymTFT. In this section we recap the introduction of \cite{Apruzzi:2021nmk} on differential cohomology in order to introduce both the notation and some of the mathematical machinery we use throughout this work. For further mathematical details and implementations of differential cohomology in string/M-theory see e.g. \cite{Freed:2000ta, Hopkins:2002rd, Freed:2006mx, bar2014differential,  Hsieh:2020jpj, Apruzzi:2021nmk}.

Differential cohomology combines information about the characteristic class of the gauge bundle and the connection. The $p^{\tn{th}}$ differential cohomology group $\br H^p(M)$ of an $n$-dimensional manifold $M$ is a differential refinement of the ordinary integral cohomology group $H^p(M;\Z)$. Denote by $\Omega^p$ closed $p$-forms, and by $\Omega_\mathbb{Z}^p$ the subset of those with  integral periods. 
The differential cohomology class takes part in the commutative diagram, whose diagonals are all short exact sequences \footnote{An alternative way of describing differential cohomology is as follows. Differential cohomology is useful to describe the non-trivial topological structure of higher-form gauge fields. A representative $\br A$ of a class $[\br A ] \in \br H^p(M)$ is specified by a tuple\cite{Hsieh:2020jpj}
\be
\br A = (N,A,F) \,.
\ee
Here $F$ is the field strength and is a closed $(p+1)$-form. $A$ and $N$ are maps from $C_p(M)$, the space of $p$-chains, to $\mathbb{R}$ and $\Z$ respectively. They encode holonomies and the non-trivial interplay between these holonomies and the field strength. See \cite{Hsieh:2020jpj} for more information on differential cohomology phrased in this way.
}:
\begin{equation}    \label{diamond}
	\begin{tikzcd}[row sep={1.732050808cm,between origins},column sep={2cm,between origins}]
		{} & {} & {} & \text{Tor} H^p(M; \mathbb{Z}) \arrow[dr] & {} & {} & {} \\ 
		{} & {} & H^{p-1}(M;\mathbb R / \mathbb Z) \arrow[rr,"-\beta"] \arrow[dr, hookrightarrow, "i" ] \arrow[ur] & {} & H^{p}(M;\mathbb Z) \arrow[dr , "\varrho"] & {} & {} \\ 
		{} & \frac{H^{p-1}(M;\mathbb R)}{H^{p-1}_{\text{Free}}(M;\mathbb{Z})} \arrow[ur] \arrow[dr  ] & {} &  \br  H ^{p}(M) 
\arrow[ur, twoheadrightarrow, "I" ] \arrow[dr, twoheadrightarrow, "R"] & {} & H^{p}_{\text{Free}}(M ;\mathbb Z) & {} \\
		{} & {} & \displaystyle \frac{ \Omega^{p-1}(M)}{\Omega_{\mathbb Z}^{p-1}(M)} \arrow[rr,"d_{\mathbb Z}" ] \arrow[ur,hookrightarrow, "\tau"  ] \arrow[rd, "d"] & {} & \Omega_{\mathbb Z}^{p}(M) \arrow[ur, "r"] & {} & {}\\ 
		{} & {} & {} & d\Omega^{p-1}(M) \arrow[ur] & {} & {} & {} 
	\end{tikzcd}
\end{equation}
Differential cohomology is endowed with a bilinear product 
\be 
\star\,: \br H^p(M) \times \br H^q(M) \rightarrow \br H^{p+q}(M)\,,
\ee 
with the properties 
\be 
\br a \star \br b= (-1)^{pq} \br b \star \br a\,, \qquad I(\br a \star \br b)= I(\br a) \smile I(\br b)\,, \qquad R(
\br a \star \br b) = R(\br a) \wedge R(\br b)\,,
\ee 
for $\br a \in \br H^p(M)$ and $\br b \in \br H^q(M)$.
It has two non-trivial integration maps, namely:
\begin{itemize}
\item the {\it primary invariant} of a differential cohomology class of degree $n=\tn{dim}(M)$
\be 
\label{eq:priminv}
\int_M \br a=\int_M I(\br a)
= \int_M R(\br a) \in \Z\,, \qquad \br a \in \br H^n(M)\,,
\ee 
\item the {\it secondary invariant} of a differential cohomology class of degree $n+1$ (see e.g. \cite{Freed:2006yc})
\be 
\label{eq:secinvdef}
\int_M \br a=\int_M w \ \tn{mod} \ 1= \int_M u \in \R/\Z\,, \qquad \br a \in \br H^{n+1}(M)\,,
\ee 
with 
\be 
\tau(w)=\br a\,,  \quad w \in \frac{\Omega^n(M)}{\Omega^n_{\Z}(M)}\,, \qquad \tn{and} \qquad \br a=i(u)\,, \quad u \in H^n(M;\R/\Z)\,.
\ee 
\end{itemize}
In the setting of M-theory, we can write the action \eqref{eq:CSaction} as the secondary invariant of a class $\br I_{12} \in \br H^{12}(\cM_{11})$,
\be
\label{eq:defStop}
\frac{S}{2\pi}=\int_{\cM_{11}}\br I_{12} \ \text{mod} \ 1\,,
\ee 
where 
\be 
\br I_{12}=- \frac{1}{6} \br G_4 \star \br G_4 \star \br G_4-\br G_4 \star \br X_8\,,
\ee 
with $\br G_4 \in \br H^4(\cM_{11})$ and $\br X_8 \in \br H^8(\cM_{11})$.

For a given 7-manifold $Y_7$, the generators of $H^p(Y_7;\Z)$, $p=0,\dots,7$ are denoted as follows:
\begin{itemize}
	\item free generators of $H^p(Y_7;\Z)$: $r(\omega_p^i) \equiv v_p^i$, $i=1,\dots,b^p(Y_7)$ with $\omega_p^i \in \Omega^p_\Z(Y_7)$,
	\item torsion generators of $H^p(Y_7;\Z)$: $t_p^\alpha$, $\alpha \in A_p$ for some set of superscripts $A_p$.
\end{itemize}
For each torsion generator, there exists a minimal positive number $\ell^\alpha_p \in \mathbb{N}$, such that 
\be 
\label{eq:torsionaldegree}
\ell^\alpha_p t_p^\alpha=0\,.
\ee 
We will be particularly interested in the secondary invariant of $\br I_{12}$ on a product space, which is the compactification space of M-theory. For a set of differential cohomology classes $\br a \in H^p(\cM_4)$, $\br b \in H^q(Y_7)$ with $p+q=12$ we have 
\be 
\label{eq:secinv}
\int_{\cM_4 \times Y_7} \br a \star \br b= \begin{cases} 	
	\lb \int_{\cM_4} u \rb \lb \int_{Y_7} R(\br b) \rb & \text{if} \quad p=5 \\
	\lb \int_{\cM_4} R(\br a) \rb \lb \int_{Y_7} s \rb & \text{if} \quad p=4
\\
	0 & \text{otherwise}
\end{cases}
\ee 
where 
\be 
i(u)= \br a\,, \qquad i(s)=\br b\,.
\ee 
Now, we can choose the differential cohomology uplifts of the torsion generators $\br t$ to be flat \cite{Apruzzi:2021nmk}
\be 
R(\br t)=0\,,
\ee 
which implies that for terms in $\br I_{12}$ involving the torsion generators $\br t$, only those with 8 internal components will contribute (i.e. with $p=4$ in \eqref{eq:secinv})

\subsection{Accounting for Torsion using Differential Cohomology}
\label{sec:diffcoho}
In this section we expand the differential refinement of $G_4$, $\br G_4 \in \br H^4(\cM_{11})$, on the product space $\cM_{11}=\cM_4 \times Y_7$ and derive the topological sector of the effective 4d supergravity theory, including torsion contributions.

We will take $\cM_4$ to be connected, so $H^0(\cM_4;\Z)=\Z$, and assume vanishing torsion  $\text{Tor}H^\bullet(\cM_4;\Z)=0$. We will furthermore assume that $Y_7$ is closed, connected and orientable, so that \cite{Hatcher:478079}
\be 
H^0(Y_7;\Z)=\Z\,, \qquad \text{Tor}H^1(Y_7;\Z)=0\,.
\ee 
Thus, we take $v_0\equiv 1$ as the generator of $H^0(Y_7)$. We can expand the ordinary cohomology class\footnote{We use $G_4$ both for the cohomology class and for the differential form representing the free part.} $G_4 \in H^4(\cM_{11};\Z)$ as
\be 
\label{eq:G4coho}
G_4= \sum_{p=0}^4 \sum_{i=1}^{b^p(Y_7)} F_{4-p}^i \smile v_p^i+\sum_{p=2}^4 \sum_{\alpha \in A_p} B_{4-p}^\alpha \smile t_p^\alpha\,.
\ee 
Here, $F^i_{q} \in H^{q}(\cM_4;\Z)$ are a set of field strengths related to $g_q^i$ in (\ref{eq:fluxexp}) by
\be 
\varrho(F_q^i)=r(g_q^i)\,,
\ee 
and $B_{q}^\alpha\in H^{q}(\cM_4;\Z)$ model a set of closed $q$-form gauge fields. In particular, let us comment on the 0-forms $F_0^i$ and $B_0^\alpha$. Due to flux quantisation (over ordinary and torsional cycles, respectively), $F_0^i$ and $B_0^\alpha$ are in fact integers. For $F_0^i$, we have $\varrho (F_0^i)=r(\cN^i)$. We will simply write
\be 
F_0^i=\cN^i\in\mb{Z}\,,
\ee 
which is background flux over internal 4-cocycles supporting the vacuum.
For $B_0^\alpha\in H^0(\mc{M}_4;\mb{Z})$, commutativity of the righthand diagram in \eqref{diamond} for $p=0$ implies the existence of a set of integers $b^\alpha \in \Omega_\Z^0$, such that $\varrho (B_0^\alpha)=r(b^\alpha)$, parametrizing the background flux over torsion 4-cocycles in $Y_7$. We write
\be 
\label{eq:bgtorsion}
B_0^\alpha=b^\alpha \in \Z\,. 
\ee 
In order to distinguish this background flux from the fluctuating fields, we will use $b^{\alpha}$ below.

The uplift to differential cohomology $\br G_4 \in \br H^4(\cM_{11})$ is performed using the surjective map $I\,: \br H^p(M) \twoheadrightarrow H^p(M;\Z)$ in \eqref{diamond}, which implies the existence of differential cohomology classes $\br F_{4-p}^i, \br B_{4-p}^\alpha \in \br H^{4-p}(\cM_4)$ and $\br v_p^i, \br t_p^\alpha \in \br H^p(Y_7)$ such that
\be 
\label{eq:diffcohofields}
F_{4-p}^i=I(\br F_{4-p}^i)\,, \qquad B_{4-p}^\alpha=I(\br B_{4-p}^\alpha)\,, \qquad  v_p^i=I(\br v_p^i)\,, \qquad t_p^\alpha=I(\br t_p^\alpha)\,.
\ee 
We therefore can write the differential cohomology uplift 
\be 
\label{eq:diffcohoG4}
\br G_4=\sum_{p=0}^4 \sum_{i=1}^{b^p(Y_7)} \br F_{4-p}^i \star \br v_p^i+\sum_{p=0}^4 \sum_{\alpha \in A_p} \br B_{4-p}^\alpha \star \br t_p^\alpha\,, 
\ee 
such that
\be 
G_4=I(\br G_4)\,.
\ee 
The map $I$ only determines $\br G_4$ up to a topologically trivial element. 
However the contribution from this element is accessible through the ordinary cohomology formulation, so we set it to zero in the following.

Dimensional reduction of the CS-term, using the expansion of $\br G_4$ in \eqref{eq:diffcohoG4} (and flatness of $\br t$) yields a significant number of potential topological couplings, which we organise by the number of continuous, respectively discrete, gauge fields (i.e. into four types of the form $F^3$, $F^2B$, $FB^2$ and $B^3$). 
Furthermore, we denote the 8-dimensional secondary invariants of $\br H^8(Y_7)$ over the internal space by\footnote{Note that the $\Lambda$'s containing $v_0^i \equiv 1$ will have one less $i,j,k$ index. E.g. we write $\Lambda^{ijk}_{0m} \equiv \Lambda^{jk}_{0m}$.}
\be 
\ba 
\label{eq:CSinvariants}
\Lambda^{ijk}_{nm} &\equiv \int_{Y_7} \br v_n^i \star \br v_{8-n-m}^j \star \br v_m^k\,,\\
\Lambda^{ij\alpha }_{nm} &\equiv \int_{Y_7} \br v_n^i \star \br v_{8-n-m}^j \star \br t_m^\alpha \,,\\
\Lambda^{i\alpha \beta }_{nm} &\equiv \int_{Y_7} \br v_n^i \star \br t_{8-n-m}^\alpha \star \br t_m^\beta\,,\\
\Lambda^{\alpha \beta \gamma}_{nm} &\equiv \int_{Y_7} \br t_n^\alpha  \star \br t_{8-n-m}^\beta \star \br t_m^\gamma\,.
\ea 
\ee 
For the $F^3$ component we obtain
\be
\ba 
\label{eq:FFF}
\left. \int_{\cM_{11}}  -\frac{1}{6}\br G_4^3 \right\vert_{F^3}= 
&\sum_{ijk} \lbb -\frac{K^{ijk}}{2} \int_{\cM_4}  \br F_3^i \star \br F_1^j \star \br F_1^k+ \frac{\cK^{ijk}}{2} \int_{\cM_4} \br F_2^i \star \br F_2^j \star \br F_1^k+ \mathbb{K}^{ijk} \cN^i \int_{\cM_4}  \br F_2^j \star \br F_3^k \right. \\
& \hspace{.75cm}+ \left. \frac{\Lambda^{ijk}_{23}}{2} \int_{\cM_4} \br F_2^i \star \br F_1^j \star \br F_1^k +\frac{\Lambda^{ijk}_{24}}{2} \cN^k \int_{\cM_4} \br F_2^i \star \br F_2^j + \frac{\Lambda^{ijk}_{14}}{2} \cN^k \int_{\cM_4} \br F_3^i  \star \br F_1^j \rbb \\
+&\sum_{ij} \lbb \mathfrak{K}^{ij} \cN^i \int_{\cM_4} \br F_1^j \star \br F_4  - \frac{\Lambda^{ij}_{40}}{2} \cN^i \cN^j \int_{\cM_4} \br F_4 \rbb \,.
\ea 
\ee 
Here we notice that the four terms with primary invariants on $Y_7$ are precisely those captured by the ordinary cohomology reduction (and we therefore use the previous notation for the coefficients). Using the definition of the primary invariant \eqref{eq:priminv} we have for $n,m=0,\dots,4$ and $3 \leq n+m \leq 7$
\be 
\ba 
\int_{Y_7} \br v_n \star \br v_{7-n-m} \star \br v_m
&=\int_{Y_7} R( \br v_n) \wedge R(\br v_{7-n-m}) \wedge R(\br v_m)\\
&=\int_{Y_7} \omega_n \wedge \omega_{7-n-m} \wedge \omega_m\,.
\ea 
\ee 
By comparison with \eqref{eq:int} we conclude that
\be 
\ba 
\label{eq:intdiffcoho}
K^{ijk}&= \int_{Y_7} \br v_1^i \star \br v_3^j \star \br v_3^k\,, \qquad &&\mathcal{K}^{ijk} = \int_{Y_7} \br v_2^i  \star \br v_2^j \star \br v_3^k\,, \\
\mathfrak{K}^{ij}&=\int_{Y_7} \br v_4^i \star \br v_3^j\,, \qquad &&\mathbb{K}^{ijk} = \int_{Y_7}  \br v_4^i \star \br v_2^j \star \br v_1^k\,.
\ea 
\ee 
Furthermore, using \eqref{eq:secinvdef} we have (again, for $n,m=0,\dots,4$ and $3 \leq n+m \leq 7$)
\be 
\int_{\cM_4} \br F_{4-n} \star \br F_{n+m-3} \star \br F_{4-m}=\int_{\cM_4} w_{nm} \ \text{mod} \ 1\,, \qquad w_{nm} \in \frac{\Omega^4(\cM_4)}{\Omega^4_\Z(\cM_4)}\,,
\ee 
where 
\be 
d_\Z w_{nm}=R(\br F_{4-n} \star \br F_{n+m-3} \star \br F_{4-m})=g_{4-n} \wedge g_{n+m-3} \wedge g_{4-m}\,.
\ee 
From this we see that \eqref{eq:FFF} reproduces all the couplings from the CS-term in $I_5$ \eqref{eq:I5Free}. The other four terms are new compared to the ordinary cohomology reduction. The $F^2B$ contribution is 
\be
\ba 
\label{eq:FFB}
\left. \int_{\cM_{11}}-\frac{1}{6}\br G_4^3 \right\vert_{F^2B}= &\sum_{ij \alpha} \lbb  \Lambda^{ij\alpha}_{33} \int_{\cM_4} \br F_1^i \star \br F_2^j \star \br B_1^\alpha + \Lambda^{ij\alpha}_{43} \cN^i \int_{\cM_4}  \br F_3^j \star \br B_1^\alpha \right. \\
& \hspace{3.5mm} - \Lambda^{ij\alpha}_{42} \cN^i \int_{\cM_4}  \br F_2^j \star \br B_2^\alpha +\Lambda^{ij\alpha}_{34} b^\alpha \int_{\cM_4} \br F_1^i \star \br F_3^j\\
& \hspace{3.5mm} -\frac{\Lambda^{ij\alpha}_{24}}{2} b^\alpha \int_{\cM_4} \br F_2^i \star \br F_2^j + \left. \frac{\Lambda^{ij\alpha}_{32}}{2} \int_{\cM_4} \br F_1^i \star \br F_1^j \star \br B_2^\alpha \rbb -\sum_{i \alpha} \Lambda^{i\alpha}_{44} \cN^i b^\alpha \int_{\cM_4} \br F_4 \,.
\ea 
\ee 
Finally, the $FB^2$ and $B^3$ terms are respectively
\be
\ba
\label{eq:FBB}
\left. \int_{\cM_{11}}-\frac{1}{6}\br G_4^3 \right\vert_{FB^2}= &\sum_{i \alpha \beta} \lbb \Lambda^{i\alpha \beta}_{32} \int_{\cM_4} \br F_1^i \star \br B_1^\alpha \star \br B_2^\beta + \frac{\Lambda^{i\alpha \beta}_{23}}{2} \int_{\cM_4} \br F_2^i \star \br B_1^\alpha \star \br B_1^\beta \right. \\
& \hspace{3.5mm}- \Lambda^{i\alpha \beta}_{22} b^\alpha \int_{\cM_4} \br F_2^i \star \br B_2^\beta + \Lambda^{i\alpha \beta}_{13} b^\alpha \int_{\cM_4} \br F_3^i \star \br B_1^\beta \\
& \hspace{3.5mm} - \left. \frac{\Lambda^{i\alpha \beta}_{42}}{2} \cN^i \int_{\cM_4}  \br B_2^\alpha \star \br B_2^\beta \rbb
- \sum_{\alpha \beta} \frac{\Lambda^{\alpha \beta}_{04}}{2} b^\alpha b^\beta \int_{\cM_4} \br F_4 \,,
\ea 
\ee 
and
\be
\ba 
\label{eq:BBB}
\left. \int_{\cM_{11}} -\frac{1}{6}\br G_4^3 \right\vert_{B^3}= \sum_{\alpha \beta \gamma } & \lbb \frac{\Lambda^{\alpha \beta \gamma}_{23}}{2} \int_{\cM_4} \br B_2^\alpha \star \br B_1^\beta \star \br B_1^\gamma  - \frac{\Lambda^{\alpha \beta \gamma}_{24}}{2}  b^\gamma \int_{\cM_4} \br B_2^\alpha \star \br B_2^\beta \rbb \,.
\ea 
\ee 
Finally, we wish to account for the higher derivative contribution from the M-theory effective action given by $\int_{\cM_{11}} C_3\wedge X_8$ with $X_8 \in H^8(\cM_{11};\Z)$ in \eqref{eq:X8}. We can promote $X_8$ (equivalently, the Pontryagin classes) to a differential cohomology class $\br X_8 \in \br H^8(\cM_{11})$ as described in \cite{bunke2013differential}. We have
\be
\br X_8 = - \frac{1}{96} \br p_1(T\cM_4) \star \br p_1(TY_7)\,.
\ee
Then 
\be\ba 
\int_{\cM_{11}}-\br G_4 \star \br X_8 &=  \sum_\alpha \lbb \frac{1}{96} \int_{Y_7} \br t_4^\alpha \star \br p_1(T Y_7) \rbb \int_{\cM_4} b^\alpha \br p_1(T\cM_4) \\
&+ \sum_i \lbb \frac{1}{96} \int_{Y_7} \br v_3^i \star \br p_1(T Y_7)\rbb \int_{\cM_4} \br F_1^i \star \br p_1(T \cM_4) \\
&+ \sum_i \lbb \frac{1}{96} \int_{Y_7} \br v_4^i \star \br p_1(T Y_7)\rbb \int_{\cM_4} \cN^i \br p_1(T \cM_4)\,.
\ea\ee
The second term again reproduces what we found using the ordinary cohomology reduction. For the remainder of this work we ignore such contributions. Notice that the first and third terms above contain no dynamical fields. The second term may in principle contribute non-trivially, but for all examples we consider these terms are absent.

\paragraph{Application: Holographic AdS$_4$ Backgrounds. }
We now turn to  \linebreak AdS$_4$/CFT$_3$ holographic setups, where the supergravity background is supported by $\cL$ units of $G_4$ background flux over AdS$_4$ and the internal space has torsion cycles. In this case the background flux that we have parametrized by $\cN^i$ in the above will not be turned on. In addition, all the examples we consider satisfy
\be
H^1(Y_7;\Z)=0\,, \qquad H^3(Y_7;\Z)=0\,,
\ee 
for which the topological action in \eqref{eq:defStop} simplifies significantly to
\be
\ba
\label{eq:topactionAdS4}
\frac{\Stop}{2\pi}=-& \sum_{ij\alpha} \frac{\Lambda^{ij\alpha}_{24}}{2} b^\alpha \int_{\cM_4} \br F_2^i \star \br F_2^j- \sum_{i\alpha \beta}\Lambda^{i\alpha \beta}_{22} b^\alpha \int_{\cM_4} \br F_2^i \star \br B_2^\beta \\
- & \sum_{\alpha \beta} \frac{\Lambda^{\alpha \beta}_{04}}{2} b^\alpha b^\beta \int_{\cM_4} \br F_4 - \sum_{\alpha \beta \gamma} \frac{\Lambda^{\alpha \beta \gamma}_{24}}{2} b^\gamma \int_{\cM_4} \br B_2^\alpha \star \br B_2^\beta  \,.
\ea 
\ee 
We briefly comment on the roles of each term in the above expression. The BF-term $b^\alpha \int_{\cM_4} \br F_2^i \star \br B_2^\beta=b^\alpha\int_{\mc{M}_4}B_2^\beta\smile F_2^i$ encodes non-commutativity of certain extended operators and enforces the requirement to pick a polarization in order to obtain an absolute QFT. After picking a polarization, in certain circumstances terms of this type can correspond to a mixed 't Hooft anomaly polynomial between a discrete 1-form symmetry $\mb{Z}_{\ell_\beta}$ with 2-form background gauge field $B_2^\beta$ and a $U(1)$ 0-form symmetry with field stength $F_2^i$. The BB-term $b^\gamma \int_{\cM_4} \br B_2^\alpha \star \br B_2^\beta=b^\gamma\int_{\mc{M}_4}B_2^\alpha\smile B_2^\beta$ is a 
't Hooft anomaly for the discrete 1-form symmetries $\mb{Z}_{\ell_\alpha}$ and $\mb{Z}_{\ell_\beta}$. Note that the presence of discrete background flux $b^\alpha \neq 0$ for some $\alpha$ is essential for the existence of the anomalies. We will not discuss the physical effects of the $\theta$-term $b^\alpha \int_{\cM_4} \br F_2^i \star \br F_2^j$ or $b^\alpha b^\beta \int_{\cM_4} \br F_4$ in this paper\footnote{Such terms in the anomaly polynomial are gauge invariant by themselves, and they do not change the equation of motion for the bulk gauge fields in the AdS/CFT interpretation.}.

\section{SymTFT Coefficients from Geometry}
\label{sec:symtftforSE7manifolds}
A crucial aspect of the above analysis is the coefficients $\Lambda$. Clearly, the numerical value of these $\int_{Y_7}$ integrals is important: a value of zero implies the absence of a particular term in the anomaly polynomial, whilst a non-zero coefficient contains physical information. In this section we determine these explicitly in the case of toric Calabi-Yau 4-folds. 

\subsection{SymTFT Coefficients from Intersection Theory}
\label{sec:coeff}

The coefficients of the 4d topological action resulting from \eqref{eq:FFF}, \eqref{eq:FFB}-\eqref{eq:BBB} are given by the primary/secondary invariants of elements of $\br H^p(Y_7)$ with $p=7,8$ over $Y_7$. In the case where the integrand is an element of $\br H^7(Y_7)$ we showed that these are simply the intersection numbers \eqref{eq:intdiffcoho} that we also obtain from the ordinary cohomology reduction.
On the other hand, when the integrand is an element of $\br H^8(Y_7)$ as in \eqref{eq:CSinvariants} there is no analogue in ordinary cohomology. Still, we would like a convenient way to evaluate these coefficients, which, it turns out, can be accessed by considering a space $\cX_8$ of which $Y_7$ is the boundary. In holography this notion is quite natural since the duality is precisely between supergravity compactified on $Y_7$ and branes probing the tip of the cone over the compactification space, i.e. we can take $\cX_8=\cC(Y_7)$ to be this cone. It should be clear that the $\Lambda$'s in \eqref{eq:CSinvariants} are defined purely in terms of the geometry of $Y_7$, and we resort to the space $\cX_8$ only for computational convenience. 

In this section we present an extension to the arguments of section 3.3 in \cite{Apruzzi:2021nmk}, where the coefficients (\ref{eq:CSinvariants}) are derived from an intersection number computation on the resolved space $\widetilde{\cX_8}$. In the geometric engineering set-up, we will assume that $\widetilde{\cX_8}$ is a non-compact Calabi-Yau 4-fold. However, the Calabi-Yau condition can be relaxed in the holography setups, such as the ABJ(M) theories in section~\ref{sec:ABJM}.

We will make use of the long exact sequence
\be 
\label{eq:longseqhom}
\dots \rightarrow H_p (\widetilde{\cX_8};\Z) \rightarrow H_p (\wcX_8,Y_7;\Z) \rightarrow H_{p-1}(Y_7;\Z) \rightarrow H_{p-1}(\wcX_8;\Z) \rightarrow \dots\,.
\ee 
Note in particular that elements of $H_p (\wcX_8;\Z)$ are {\it compact} $p$-cycles in $\wcX_8$ and elements of $H_p (\wcX_8,Y_7;\Z)$ are {\it non-compact} $p$-cycles in $\wcX_8$. We assume that there are no compact $(7-n)$-cycles in $\wcX_8$
\be
\label{eq:assumcoho}
H_{7-n}(\wcX_8;\Z)=0\,,
\ee 
for a specific $n\in \{0,\dots,6\}$.

This implies that any $(7-n)$-cycle in $Y_7$ can be realised as the boundary of an $(8-n)$-chain in $\wcX_8$. In the examples we consider, $\wcX_8$ is a non-compact toric 4-fold, and these have no non-trivial odd-dimensional cycles, $H_{2k-1}(\wcX_8;\mb{Z})\equiv 0$ for $k\in \mathbb{N}$. 
Using Poincar\'e duality in $Y_7$, from \eqref{eq:longseqhom} we find
\be 
\label{eq:seqhom}
H_{8-n} (\wcX_8;\Z) \overset{A}{\rightarrow} H_{8-n} (\wcX_8,Y_7;\Z) \overset{f}{\rightarrow} H^n(Y_7;\Z) \rightarrow 0\,.
\ee 
Since $f$ is surjective, we conclude that {\it every $n$-cocycle in $Y_7$ can be mapped to a non-compact $(8-n)$-cycle D in $\wcX_8$}. Furthermore, a torsion class $t_n \in H^n(Y_7;\Z)$ satisfies
\be
\ell t_n=0\,,
\ee 
for some (minimal) $\ell \in \mathbb{N}$. Then exactness of \eqref{eq:seqhom} implies that there exists a compact $(8-n)$-cycle $Z \in H_{8-n}(\wcX_8;\Z)$ such that
\be 
A(Z)=\ell T\,, \qquad f(T)=t_n\,,
\ee 
which we use to { \it map a torsion class $t_n \in H^n(Y_7;\Z)$ to a compact $(8-n)$-cycle $Z$ in $\wcX_8$.}
Taking $A$ to be the intersection pairing in $\wcX_8$, the coefficients \eqref{eq:CSinvariants} of the SymTFT can be computed as follows.

We associate to $t_n^\alpha$ of torsional degree $\ell_n^\alpha$ a compact $(8-n)$-cycle $Z_\alpha^{8-n}$ in $\wcX_8$, and to $v_m^i$ a non-compact $(8-m)$-cycle $D^{8-m}_i$ in $\wcX_8$. The coefficients are then given by
\be 
\ba 
\label{eq:CSinvintersect}
\Lambda^{ijk}_{nm} &=\lbb D_i^{8-n} \cdot D_j^{n+m} \cdot D_k^{8-m} \rbb_{\text{mod} \ 1}\,,\\
\Lambda^{ij\alpha }_{nm} &= \lbb \frac{D_i^{8-n} \cdot D_j^{n+m} \cdot Z_\alpha^{8-m}}{\ell_m^\alpha} \rbb_{\text{mod} \ 1}\,,\\
\Lambda^{i\alpha \beta }_{nm} &=\lbb \frac{D_i^{8-n} \cdot Z_\alpha^{n+m} \cdot Z_\beta^{8-m}}{\ell_{8-n-m}^\alpha \ell_m^\beta} \rbb_{\text{mod} \ 1}\,,\\
\Lambda^{\alpha \beta \gamma}_{nm} &=\lbb \frac{Z_\alpha^{8-n} \cdot Z_\beta^{n+m} \cdot Z_\gamma^{8-m}}{\ell^\alpha_n \ell^\beta_{8-n-m} \ell_m^\gamma} \rbb_{\text{mod} \ 1}\,,
\ea 
\ee 
where $\cdot$ denotes intersections in $\wcX_8=\mc{C}(Y_7)$.
 
We must take extra care with the terms in the 4d effective action which come with a factor of a half. That is, the relevant object to compute is not $\Lambda$ but rather $\Omega \equiv \Lambda/2$. However multiplication by $1/2$ is not a well-defined operation due to the mod 1 in \eqref{eq:CSinvintersect}. Instead, using the approach by Gordon and Litherland \cite{GordonLitherland}, the refinement by a factor of $1/2$ is accounted for by computing
\be 
\ba\label{eq:coefficients} 
\Omega^{ij}_{n}&=\half \int_{Y_7} \br v_n^i \star \br v_n^i \star \br v_{8-2n}^j=\lbb \frac{D_i^{8-n} \cdot D_i^{8-n} \cdot D_j^{2n}}{2} \rbb_{\text{mod} \ 1}\,,\\
\Omega^{i\alpha }_{n}&= \half \int_{Y_7} \br v_n^i \star \br v_n^i \star \br t_{8-2n}^\alpha=\lbb \frac{D_i^{8-n} \cdot D_i^{8-n} \cdot Z_\alpha^{2n}}{2\ell_{8-2n}^\alpha} \rbb_{\text{mod} \ 1} \,,\\
\Omega^{i\alpha}_{n}&= \half \int_{Y_7} \br v_{8-2n}^i \star \br t_{n}^\alpha \star \br t_n^\alpha=\lbb \frac{D_i^{2n} \cdot Z_\alpha^{8-n} \cdot Z_\alpha^{8-n}}{2 (\ell_n^\alpha)^2} \rbb_{\text{mod} \ 1}\,,\\
\Omega^{\alpha \beta}_{n}&= \half \int_{Y_7} \br t_n^\alpha  \star \br t_n^\alpha \star \br t_{8-2n}^\beta=\lbb \frac{Z_\alpha^{8-n} \cdot Z_\alpha^{8-n} \cdot Z_\beta^{2n}}{2 (\ell^\alpha_n)^2 \ell^\beta_{8-2n}} \rbb_{\text{mod} \ 1}\,,
\ea
\ee
which are $\R/\Z$-valued quantities.

\subsection{Intersection Numbers of Toric 4-Folds}
\label{sec:intersectionoftoriccy4}
The above subsection explained that the computation of the SymTFT coefficients reduces to a computation of intersection numbers in $\mc{X}_8$. In this section we focus on \textit{toric} 4-folds. It will become apparent in later sections that the non-trivial coefficients we are particularly interested in are those involving $\br t_2^\alpha, \br t_4^\alpha$ and $\br v_2^i$. The key identifications to make are therefore the \textit{compact} divisor $Z^6$ corresponding to $\br t_2^\alpha$, the \textit{compact} 4-cycles $Z^4_{\beta}$ corresponding to $\br t_4^{\beta}$ and the \textit{non-compact} divisor $D^6$ corresponding to $\br v_2^i$. 
We will address the identifications in turn. First however, we introduce the technology required to compute intersection numbers of toric 4-folds.

\paragraph{Quadruple Toric Intersections. } All non-zero integrals of the type we wish to consider reduce to a sum of quadruple intersection of toric divisors $T_i$ in the Calabi-Yau
\be
T_i \cdot T_j \cdot T_k \cdot T_l \,.
\ee
We begin with a toric fourfold $\wcX_8$, described by a toric diagram with set of rays $\{v_i \}$,
\be
v_i=(v_i^x,v_i^y,v_i^z,v_i^w)\,.
\ee
Each ray corresponds to a toric divisor $v_i \leftrightarrow T_i$, amongst which there exists a set of linear relations
\be
\label{eq:linear-equiv}
\sum_i v_i^x T_i = 0\,, \quad \sum_i v_i^y T_i = 0 \, \quad \sum_i v_i^z T_i = 0 \,, \quad\sum_i v_i^w T_i = 0 \,.
\ee
Furthermore, we triangulate the toric diagram with a set of 4d cones
\be
\{v_a v_b v_c v_d \}\,,
\ee
which restrict the non-zero quadruple  intersections in the following way. The intersection of four distinct toric divisors is given by the volume bounded by the rays (we denote this region $V_{ijkl}$)
\be
\label{quad-int-4d}
T_i \cdot T_j \cdot T_k \cdot T_l = \frac{1}{\text{vol}(V_{ijkl})} \,, \quad i\neq j \neq k \neq l \,.
\ee
The quadruple intersection numbers involving self-intersections can be computed using (\ref{quad-int-4d}) and the linear equivalence relations \eqref{eq:linear-equiv} \footnote{In order to perform such calculations in practise, a computer code is necessary for larger toric diagrams. Assuming we have computed $T_i \cdot T_j \cdot T_k \cdot T_l $ for all distinct $i \neq j \neq k \neq l$ we can compute the following intersections in turn
\begin{enumerate}
\item $T_i \cdot T_i \cdot T_j \cdot T_k \, \quad i \neq j \neq k $
\item $T_i \cdot T_i \cdot T_i \cdot T_k \,, \quad i \neq k$ and $T_i \cdot T_i \cdot T_k \cdot T_k \,, \quad i \neq k$ 
\item $T_i \cdot T_i \cdot T_i \cdot T_i $ \,.
\end{enumerate}
In each step we use the intersections computed in the step prior.}\footnote{Note that in the cases of a non-compact toric 4-fold, the quadruple intersection numbers only involving non-compact divisors are usually not well-defined. Nonetheless, we do not encounter this issue as we only use the intersection numbers which involve at least one compact divisor, see  analogous computations in the case of CY3 \cite{Xie:2017pfl,Eckhard:2020jyr}.}. 

Now we consider the case of a toric Calabi-Yau 4-fold $\wcX_8$, such that the boundary 7-manifold $Y_7$ is Sasaki-Einstein. The Calabi-Yau condition forces the rays $\{v_i\}$ to lie in a plane. We enforce this in coordinates by choosing the fourth coordinate of all rays to be 1
\be
v_i = (\underline{v_i},1)=(v_i^x,v_i^y,v_i^z,1) \,.
\ee
For a 4d cone $v_i v_k v_k v_l$, the volume of $V_{ijkl}$ takes the form of 
\be
\text{vol}(V_{ijkl}) = \text{det} \begin{pmatrix} \underline{v}_j-\underline{v}_i & \underline{v}_k-\underline{v}_i & \underline{v}_l-\underline{v}_i
\end{pmatrix} \,.
\ee

\subsection{Differential Cohomology Generators and Toric Divisors}
\label{sec:mapbetween}
\paragraph{$\br t_2$ generators. } 

From the set of divisors $\{T_i\}$ and the linear relations between them, we can obtain a set of \textit{linearly indepent} divisors. We denote them $C_a\,, D_a $ for compact and non-compact respectively. From these, we can construct a basis of compact curves
\be
\{\cN_k\}=\{C_a \cdot D_b \cdot D_c\ ,\ C_a\cdot C_b\cdot D_c\ ,\ C_a\cdot C_b\cdot C_c\}\,.
\ee
In general these curves are not linearly independent. For example, for the case $Y^{p,k}(\bC \P^2)$, the curves $C_a\cdot D_b\cdot D_c$ already form a complete basis of compact curves.

In order to obtain the central divisors $Z^6$, we compute the SNF of the intersection matrix $ \cN_k \cdot C_a$
\be
\text{SNF}(\cN_k \cdot C_a) = \begin{pmatrix} \Gamma_1 & 0 & \dots & 0 \\
0 & \Gamma_2 & \dots & 0\\
\vdots & \vdots & \ddots &\vdots \\
0 & 0 & \dots & \Gamma_n \\
0 & 0 & \dots & 0\\
\vdots & \vdots & \ddots & \vdots\\
0 & 0 & \dots & 0
\end{pmatrix} =  A \cdot (\cN_k \cdot C_a) \cdot B\,,
\ee
where $A$ and $B$ are matrices and $\Gamma_\alpha$ are a set of integers. The group 
\be
\Gamma = \oplus_{\alpha=1}^n \Z_{\Gamma_\alpha} \,,
\ee
is generated by a set of linear combinations of divisors given by the matrix $B$.

Thus the change of basis matrices used in the SNF procedure can be used to find explicit expressions for the compact divisor dual to $\br t_2$ in terms of the basis elements $C_a$. For each $\Gamma_\alpha>1$, there is a differential cohomology class $\br t_2^\alpha$ with torsion degree $\Gamma_\alpha$. Furthermore, it is clear that the group $\Gamma$ is in fact equal to the 1-form symmetry
\be
\Gamma^{(1)} = \Gamma \,.
\ee
In particular, we can read off the generators as follows. The linear combination of divisors generating the factor $\Gamma_\alpha$ is given by
\be
Z^6_\alpha = \sum_i B_{i\alpha} C_i \,.
\ee

\paragraph{$\br t_4$ generators. } Analogously to the above procedure, we wish to identify the appropriate linear combination of 4-cycles $Z^4_{\alpha}$ dual to $\br t_4^{\alpha}$. We can construct a basis for compact 4-cycles by $\{\mc{S}_k\}=\{C_a \cdot D_b,\ C_a\cdot C_b \}$ . Once again we take the SNF of the intersection matrix
\be
\text{SNF}\{ \mc{S}_j \cdot \mc{S}_k\} =\text{diag}(\Gamma'_1,\Gamma'_2,\dots) = A' \cdot \{ \mc{S}_j \cdot \mc{S}_k\} \cdot B'\,.
\ee
We derive that the group
\be
\text{Tor}H^4 = \oplus_{\alpha} \Gamma'_\alpha \,,
\ee
is generated by the linear combinations
\be
Z^4_{\alpha} = \sum_i B'_{i\alpha} \mc{S}_i \,.
\ee
We observe that a consistent choice must be made of ordering of columns in the SNF process when two different $Y^{p,k}$ models are compared. A change of basis of the matrix corresponds to choosing different diagonal combinations of symmetries inside the group $\oplus_i \Gamma'_i$.

\paragraph{$\br v_2$ generators. } In general, the number of independent $\br v_2$ generators is equal to $b_3(Y_7)$. One can pick any of the $b_3(Y_7)$ linearly independent non-compact divisors as $\br v_2$ generators, and they will give the same physical results.

\section{SymTFT for Holography:  ABJ(M)}
\label{sec:ABJM}

We now employ the geometric tools developed in the previous sections to derive from M-theory the global structure and higher-form symmetries for the 3d $\cN=6$ $U(N+b)_k \times U(N)_{-k}$ ABJ(M) theories \cite{Aharony:2008ug,Aharony:2008gk}. In the brane picture, the theories arise on $N$ M2-branes probing a $\bC^4/\Z_k$ singularity, together with $b$ fractional M2-branes localised at the orbifold singularity. The 11d supergravity dual is $\ads{4} \times S^7/\Z_k$ with $N$ units of $G_4$ flux over $\ads{4}$ and $b$ units of torsion flux. The 7-manifold $S^7/\mb{Z}_k$ is generally a tri-Sasakian manifold, and it is Sasaki-Einstein only when $k=4$. All 3d $\cN=6$ ABJM type theories were classified, up to discrete quotients of the gauge group, in \cite{Schnabl:2008wj}, which was subsequently extended to account for all global forms in \cite{Tachikawa:2019dvq}. 
In \cite{Bergman:2020ifi} it was shown how to realise different global forms of the gauge group holographically from type IIA supergravity (in the absence of background torsion, i.e. for $b=0$) in the regime $k \ll N \ll k^5$ where the M-theory circle is small. We reproduce these results, taking the perspective of 11d supergravity, where the technology presented in previous sections is crucial to understand the geometric origin of the symmetry background fields. Moreover, with $b$ turned on, we determine a 't Hooft anomaly for the 1-form symmetry 
\be 
\label{eq:toranom1}
-\frac{b}{2k} \int_{\ads{4}} B_2 \smile B_2\,.
\ee 
We derive the anomaly from torsional geometric data, and match with field theory results \cite{Tachikawa:2019dvq}. The SymTFT, computed using differential cohomology, is precisely the tool suited to pick up such a torsional effect.

\subsection{Global Form of the Gauge Group}
\label{sec:ABJMglobalform}
The global form of the gauge group is associated with a choice of boundary conditions for the gauge fields of the 4d bulk theory \cite{Aharony:1998qu,Witten:1998wy,GarciaEtxebarria:2019caf,
Morrison:2020ool,Albertini:2020mdx,Apruzzi:2021nmk}. This choice is constrained by the fact that, in the presence of torsion in the homology of the internal space $Y_7=S^7/\Z_k$, the $G_4$ and $G_7$ fluxes do not commute at the boundary \cite{Freed:2006ya,Freed:2006yc}. 
In the SymTFT this non-commutativity of fluxes shows up as a set of BF-couplings that constrain the consistent set of boundary conditions which can be imposed on the participating 4d gauge fields.
The BF-terms arise from the differential cohomology reduction of the kinetic term in the 11d supergravity action, see \cite{saghar} for a derivation\footnote{We thank I\~{n}aki Garc\'ia Etxebarria and Saghar Sophie Hosseini for explaining this to us, and refer the reader to their upcoming work \cite{saghar} for more details.}. In the following however, we will instead take an operator perspective and derive the commutation relation.

In holography the procedure for choosing asymptotic values of the fields at the conformal boundary of $\cM_{11}=\cM_4 \times Y_7$ is to quantize the theory on $\cM_{11}=\R_t \times \cM_{10}$, by identifying the radial direction with time, and choosing a state in the Hilbert space of $\cM_{10}=M_3 \times Y_7$, where $M_3$ is the conformal boundary of $\cM_4$ at infinity \cite{Aharony:1998qu,Witten:1998wy}.
Consider the operators $\Phi(\cT_3)$ and $\Phi(\cT_6)$ which detect the periods of the M-theory gauge potential $C_3$ and the electric-magnetic dual potential $C_6$ over 3- and 6-cycles $\cT_3, \cT_6$ defining torsion homology classes in $\cM_{10}$. As shown in \cite{Freed:2006ya,Freed:2006yc}, these operators pick up a phase under commutation
\be
\Phi(\cT_3)\Phi(\cT_6)=\Phi(\cT_6)\Phi(\cT_3)e^{2 \pi i \mathsf{L}(\cT_3,\cT_6)}\,,
\ee 
where $\mathsf{L}$ is the linking pairing. The homology of $S^7/\Z_k$ is
\be 
H_\bullet (S^7/\Z_k;\Z)=\{ \Z, \Z_k, 0, \Z_k, 0, \Z_k, 0, \Z \}\,.
\ee 
Since we are assuming Tor$_\bullet(M_3;\Z)=0$, we can apply the K\"unneth formula to obtain
\be \ba 
\tn{Tor} H_3(\cM_{10};\Z)&=H_{2}(M_3;\Z) \otimes H_1(S^7/\Z_k;\Z) \oplus H_{0}(M_3;\Z) \otimes H_3(S^7/\Z_k;\Z)\,,\\
\tn{Tor} H_6(\cM_{10};\Z)&= H_3(M_3;\Z) \otimes H_3(S^7/\Z_k;\Z) \oplus H_{1}(M_3;\Z) \otimes H_5(S^7/\Z_k;\Z)\,.
\ea \ee 
This implies that the torsional 3- and 6-cycles of $\cM_{10}$ must be of the form
\be 
\cT_3=\Sigma_2 \times T_1\,, \qquad \cT_3=\Sigma_0 \times T_3\,,
\ee 
and 
\be 
\cT_6=\Sigma_3 \times T_3\,, \qquad \cT_6=\Sigma_1 \times T_5\,,
\ee 
where $\Sigma_p$ generates $H_p(M_3;\Z)$ and $T_q$ generates $H_q(S^7/\Z_k;\Z)=\Z_k$ for $q < 7$ odd.
Consider the expansion on cohomology of $G_4$ around the ABJ(M) background 
\be 
\label{eq:G4cohoABJM}
G_4=N \vol_{\ads{4}} + B_2 \smile t_2+ b \smile t_4\,,
\ee 
where $t_2$ and $t_4$ are both torsional generators of degree $k$. The differential cohomology uplift is
\be 
\br G_4=N \br \vol_{\ads{4}} + \br B_2 \star \br t_2+ \br b \star \br t_4\,,
\ee 
Here, $\br B_2$ represents a dynamical $\Z_k$ 2-form gauge field, whereas 
\be 
I(\br b) = b \in \Z\,,
\ee
is an integer parametrizing background flux over torsion 4-cocyles, as argued around \eqref{eq:bgtorsion}. This discrete flux is associated to $b$ M5-branes wrapping the torsion 3-cycle $H_3(S^7/\Z_k;\Z)=\Z_k$. In the ABJ paper \cite{Aharony:2008gk} it was conjectured that we must have $b \leq k$ for the superconformal $U(N+b)_k \times U(N)_{-k}$ theories to exist as unitary theories. As was also noted in \cite{Aharony:2008gk}, this restriction is consistent with the interpretation of $b$ as discrete $\Z_k$ torsion. Without imposing a relation between $G_4$ and $G_7$, we can make a corresponding cohomology expansion of the latter
\be 
\label{eq:G7cohoABJM}
G_7=N \vol_{S^7/\Z_k} + B_{3} \smile t_4+ B_{1} \smile t_6\,.
\ee 
In order to quantize on $\cM_{11}=\R_t \times \cM_{10}$, we consider a gauge as in \cite{Witten:1998wy} where the form representatives of the $B_p$ classes do not have components along the radial direction (or time direction, in terms of the quantization scheme), i.e. they can be taken to define either degree-$p$ cohomology classes in $\cM_4$, as in \eqref{eq:G4cohoABJM}, \eqref{eq:G7cohoABJM}, or in $M_3$. Then, by Poincar\'e duality, the integral homology classes $\Sigma_q$ are dual to $B_{3-q}$ in the (torsion-free) cohomology of $M_3=\R^{1,2}$. 

For the present, let us ignore the discrete background flux and study the ABJM theories whose type IIA duals were studied in \cite{Bergman:2020ifi}.
Then the non-commuting operators are $\Phi(\Sigma_2)$ and $\Phi(\Sigma_1)$, which are push-forwards of the 11d operators $\Phi(\Sigma_2 \times T_1)$ and $\Phi(\Sigma_1 \times T_5)$. Their commutation relation is determined using
\be 
\mathsf{L}(\Sigma_p \times T_q,\Sigma_{p'} \times T_{q'})=(\Sigma_p \cdot \Sigma_{p'}) \mathsf{L}_{S^7/\Z_k}(T_q,T_{q'})\,,
\ee 
with $\Sigma_p \cdot \Sigma_{p'}$ the intersection in $M_3$ and \cite{GarciaEtxebarria:2019caf}
\be 
\mathsf{L}_{S^7/\Z_k}(T_1,T_5)=\frac{1}{k}\,.
\ee 
Hence, we have
\be 
\label{eq:commABJM}
\Phi(\Sigma_2)\Phi(\Sigma_1)=\Phi(\Sigma_1)\Phi(\Sigma_2)e^{2 \pi i (\Sigma_1 \cdot \Sigma_2)/k}\,.
\ee 
If we consider the form representatives of $B_2, B_1$, and abuse notation by denoting them the same as their corresponding cohomology classes, this commutation relation is encoded in a BF-coupling 
\be 
\label{eq:BF-ABJM}
\boxed{ 
\frac{S_\tn{BF}}{2\pi}= k \int_{\ads{4}} B_2 \wedge d B_1\,.
}
\ee 
As we remarked above, this coupling can alternatively be derived by considering the differential cohomology reduction of the kinetic part of the 11d supergravity action, see \cite{saghar}.

The symmetries of the 3d field theory
are determined by imposing boundary conditions on $B_2, B_1$ consistent with the commutation relation \eqref{eq:commABJM}, or equivalently the action \eqref{eq:BF-ABJM}. Fixing $B_2$ to a background value as we approach the conformal boundary is associated with a 1-form symmetry, whereas fixing $B_1$ would furnish a background gauge field for an ordinary 0-form symmetry.

First however, we must also take into account the additional global 0-form symmetries that can arise from gauging the isometry group of the internal space.
The isometries of $S^7/\Z_k$ are $U(1) \times SU(4)_R$, with the latter realising the R-symmetry of the 3d $\cN=6$ theory. We can describe $S^7/\Z_k$ as a circle bundle over $\bC \P^3$ with metric \cite{Aharony:2008ug} 
\be 
ds^2_{S^7/\Z_k}=\frac{1}{k^2}(\d \varphi+k w)^2+ds^2_{\bC \P^3}\,,
\ee 
with $\varphi \sim \varphi +2\pi$ parametrizing the M-theory circle, and $dw=J$ with $J$ the K\"ahler form on $\bC \P^3$. The $\Z_k$ quotient is simply making the M-theory circle smaller, and $\partial_\varphi$ is generating the $U(1)$ isometry. We can gauge the isometries by lifting the 4-form flux to equivariant cohomology, which gives rise to a single 1-form gauge field $A_1$ for the $U(1)$ isometry
\be 
d\varphi \rightarrow d\varphi+A_1\,,
\ee 
and a set of 15 gauge fields for the $SU(4)_R$. We are interested in the fate of the $U(1)$ global symmetry, and whether it couples to the $B_2$, $B_1$ gauge fields in \eqref{eq:BF-ABJM}. We will answer this question by conjecturing a map to the type IIA description.

When $N \gg k^5$ the appropriate supergravity description is 11-dimensional. On the other hand, when $k \ll N \ll k^5$ the M-theory circle becomes very small (in Planck units) and the relevant description is type IIA supergravity on $\ads{4} \times \bC \P^3$ with $N$ units of $F_6$-flux over $\bC \P^3$ and $k$ units of $F_2$-flux over $\bC \P^1 \subset \bC \P^3$. (In the presence of background torsion flux $b \neq 0$ the NSNS 2-form on $\bC \P^1 \subset \bC \P^3$ has a discrete holonomy $b/k$.) 
Consider the type IIA supergravity analysis in \cite{Bergman:2020ifi}, where a topological term 
\be 
\frac{S_\tn{IIA}}{2\pi}=\int_{\ads{4}} B_\tn{NS} \wedge d(kA_\tn{D4}+N A_\tn{D0})\,,
\ee 
was identified and the consistent boundary conditions were studied in detail. Here $B_\tn{NS}$ is the NS-NS 2-form, and $A_\tn{D4}$, $A_\tn{D0}$ are $U(1)$ 1-form gauge fields that couple electrically respectively to D4-branes wrapping $\bC \P^2 \subset \bC \P^3$ and D0-branes. 
Under dimensional reduction from 11d supergravity to type IIA (see e.g. \cite{Becker:2006dvp}), the $U(1)$ gauge field $A_1$ associated with the isometry generated by the M-theory circle direction $\partial_\varphi$ gives rise to the 1-form gauge field $A_\tn{D0}$ sourced by D0-branes. The 1-form gauge field $B_1$ couples electrically to M5-branes wrapping the torsional 5-cycle, which descends to D4-branes wrapping $\bC \P^2$ coupled electrically to the 1-form $A_\tn{D4}$. Finally, $B_2$ couples electrically to M2-branes wrapping the torsional 1-cycle associated with the M-theory circle. Under dimensional reduction these M2-branes become fundamental strings coupling to the NS-NS 2-form. Therefore, we conjecture a map
\be 
A_1 \leftrightarrow A_{D0}\,, \qquad B_1 \leftrightarrow A_{D4}\,, \qquad B_2 \leftrightarrow B_\tn{NS}\,.
\ee 
Using this map implies the existence of a topological coupling between $B_2$ and $F=dA_1$ in M-theory, to which either equivariant or differential cohomology are not sensitive by themselves -- for a discussion of this see appendix \ref{sec:isometries}\footnote{We propose that a full equivariant differential cohomology treatment of the problem will give rise to an improvement of $G_4^\tn{eq}$ by a term involving the 2-form $B_2$ and the $U(1)$ field strength $F$ like
\be 
\label{eq:G4eqconjecture}
G_4^\tn{eq}=N \vol_{\ads{4}}+B_2 \wedge F + \cdots\,.
\ee 
}.
The 11d kinetic term is then
\be 
\label{eq:SkinABJM}
\boxed{
\frac{S_\tn{kin}}{2\pi}= \int_{\ads{4}} k B_2 \wedge dB_1+N B_2 \wedge F\,.
}
\ee 
The different global forms of the gauge group are realised holographically by imposing boundary conditions consistent with this BF-coupling. This part of the analysis is now completely analogous to \cite{Bergman:2020ifi}. For convenience, we here give a brief summary of one extreme possibility, namely $(U(N)_k \times U(N)_{-k})/\Z_{k}$.

Suppose we apply the conditions
\be 
A_1, B_1 \ \tn{Dirichlet}\,, \qquad B_2 \ \tn{Neumann}\,,
\ee 
which constrains the boundary values of the 1-forms to satisfy $kB_1+NA_1=0$. Hence, the 1-form background gauge field we can specify at the boundary gives rise to $\Gamma^{(0)}=U(1) \times \Z_{\tn{gcd}(N,k)}$, where the $U(1)$ is supplied by the diagonal combination $(B_1,A_1)=(p \cA, -q \cA)$, with $p \cdot \tn{gcd}(N,k)=N$ and $q \cdot \tn{gcd}(N,k)=k$, which decouples from the action. If $N=n n'$ for some integers $n, n'$ more complicated boundary conditions are possible. These realise the gauging of a  subgroup of the 1-form symmetry. 

\subsection{'t Hooft Anomaly for the 1-Form Symmetry}
\label{sec:ABJManomalysec}
We now turn on the torsion flux $b \neq 0$ and see how the theory is modified. From the differential cohomology reduction of the 11d Chern-Simons term we determine the Symmetry TFT coupling
\be
\frac{S_\text{top}}{2\pi}= 
- \Omega \int_{\ads{4}} \br B_2 \star \br B_2 \star \br b \quad \text{mod } 1 \,,
\ee
with
\be
\label{ABJM-Omega}
\ba 
\Omega&= \half \int_{S^7/ \Z_k} \br t_2 \star \br t_2 \star \br t_4 =\lbb \frac{Z^6 \cdot Z^6 \cdot Z^4}{2 k^3} \rbb_{\text{mod} \ 1}\,,
\ea \ee 
The primary invariant over $\ads{4}$ gives
\be 
\label{eq:toranom2}
\frac{S_{\tn{top}}}{2\pi}=-\Omega b \int_{\ads{4}} B_2 \smile B_2\,,
\ee 
which signals an anomaly in the $\Z_k$ 1-form symmetry of $U(N+b)_k \times U(N)_{-k}$, determined by the coefficient $\Omega$. The 4d term \eqref{eq:toranom2} constrains the possibility of gauging a $\Z_{m}$ subgroup, with $k=m m'$, of the $\Z_k$ 1-form symmetry. That is, the anomaly is only consistent with $B_2$ having periodicity $\Z_{m'}$ for $\Omega b m'^2=0$ mod $\half$ (since $B_2 \smile B_2$ is even on a spin manifold) or, equivalently,
\be 
\label{ABJM-condition}
\frac{2 \Omega k^2 b}{m^2}=0 \ \tn{mod} \ 1\,.
\ee
To compute the coefficient $\Omega$ geometrically from (\ref{ABJM-Omega}), let us consider the resolution $\wcX_8$ of $\mb{C}^4/\mb{Z}_k$. Note that the singularity $\mb{C}^4/\mb{Z}_k$ has a toric description, with the rays
\be
v_1=(1,0,0,0)\ ,\ v_2=(0,1,0,0)\ ,\ v_3=(0,0,1,0)\ ,\ v_4=(-1,-1,-1,k)
\ee
and the 4d cone $v_1 v_2 v_3 v_4$. It has a unique toric resolution  $\wcX_8=\widetilde{\mb{C}^4/\mb{Z}_k}$, where the compact exceptional divisor $C$ corresponds to the new ray $v_5=(0,0,0,1)$\footnote{Restricting to toric varieties, $\widetilde{\mb{C}^4/\mb{Z}_k}$ is the unique resolution because $v_5=(0,0,0,1)$ is the only primitive ray inside the cone $v_1 v_2 v_3 v_4$.}. The new set of 4d cones in $\widetilde{\mb{C}^4/\mb{Z}_k}$ is $\{v_1 v_2 v_3 v_5$, $v_1 v_2 v_4 v_5$, $v_1 v_3 v_4 v_5$, $v_2 v_3 v_4 v_5\}$. Denote the non-compact divisor corresponding to $v_1$ by $D$ (which is linearly equivalent to $v_2$, $v_3$ and $v_4$). We can compute the following intersection numbers
\be
\label{ABJM-int}
D^4=0\,,\quad C\cdot D^3=1\,, \quad C^2\cdot D^2=-k\,,\quad C^3\cdot D=k^2\,,\quad C^4=-k^3\,,
\ee
from which we obtain the generators $Z^4=C\cdot D$, $Z^6=C$\,. Note that the resolution is only crepant and leading to a toric Calabi-Yau fourfold when $k=4$. Nonetheless, this is not a problem since it is not necessary to define the SymTFT in a supersymmetric way and one can use any (possibly non-crepant) resolution for the purpose of computing the SymTFT action.
Hence we can plug (\ref{ABJM-int}) into (\ref{ABJM-Omega}) to get
\be
\Omega=\frac{1}{2k}\ \text{mod}\ 1\,.
\ee
Recalling the condition (\ref{ABJM-condition}), the gauging is consistent for 
\be 
\label{eq:anomABJ}
\frac{k b}{m^2}=0 \ \tn{mod} \ 1\,.
\ee 
The anomaly thus implies that gauging a $\Z_{m}$ subgroup of the 1-form symmetry of $U(N+b)_k \times U(N)_{-k}$ is consistent only for certain choices of $m$. That is, compared to the analysis at the end of the previous section, when $b$ is turned on, certain global forms of the 3d gauge group are no longer consistent. E.g. in the presence of this anomaly we can only gauge the full 1-form symmetry $m=k$, if $b/k \in \Z$.

Note that this anomaly was also determined from the field theory point of view in \cite{Tachikawa:2019dvq}, where the authors show that, the anomaly can be measured by the topological spin of a line of charge $m'$, where $k=m m'$. In \cite{Tachikawa:2019dvq} the anomaly free lines were determined to be exactly the ones satisfying \eqref{eq:anomABJ}.

\section{SymTFT for Holography: AdS$_4 \times Y^{p,k}(\mathbb{C}\mathbb{P}^2)$}
\label{sec:symtftholopart2}
We now apply the Symmetry Topological Field Theory technology to a class of holographic 3d $\cN=2$ QFTs. We study the theories living on the worldvolume of a stack of M2-branes probing the cone $\cC(Y^{p,k}(\mathbb{C}\mathbb{P}^2))$ with torsional $G_4$ flux turned on \cite{Benini:2011cma}. The latter phenomenon arises from wrapped M5-branes on the torsional elements of the third homology group of $Y^{p,k}$ (which is non-trivial). 
The purpose of this section is to put into practise the machinery developed in the preceding sections of this paper in an intricate holographic setup, where the SymTFT can be used to derive new constraints on the 3d field theory. In particular, we derive SymTFT terms via M-theory reduction and compute the relevant coefficients using the toric CY$_4$ methods explained in section \ref{sec:symtftforSE7manifolds}.

The BF-terms we obtain are
\be
\frac{S_{\tn{BF}}}{2\pi} =\int B_2 \wedge \left( Nf_2 + \text{gcd}(p,k)\widetilde{g}_2 + \Omega^{p,k}_{n_0,n_1} g_2 \right) \,,
\ee
with integral coefficients $\Omega^{p,k}_{n_0,n_1}$ which depend on $p,k$ as well as the $G_4$ torsion flux parameterized by two integers $(n_0,n_1)$. Furthermore, in many cases we derive new 1-form symmetry anomalies of the form
\be
\Omega_{BB} \int B_2 \smile B_2 \,,
\ee
for 1-form symmetry background fields $B_2$ and some coefficient $\Omega_{BB}$ which we compute.

The dual field theories described in \cite{Benini:2011cma} are subtle and furthermore not completely constrained. We discuss the matching of our results with this work in section \ref{sec:fieldtheorymatching}, and comment on how the SymTFT could be used to solve some ambiguities.

It should be noted that more generally one could consider a stack of M2-branes probing the cone $\cC(Y^{p,k}(B))$ for more generic base $B$ \cite{Closset:2012ep,Benini:2009qs}. For example, for $B = \mathbb{C}\mathbb{P}^1 \times \mathbb{C}\mathbb{P}^1$ the SymTFT is almost identical in structure, differing only in the number of $\br v_2$ generators, and therefore $\br F_2$ background fields.

\subsection{SymTFT for General $p,k$}
In this section we perform the torsional reduction detailed in section \ref{sec:SymTFT} here for 
\be
\cM_{11} = \text{AdS}_4 \times Y^{p,k}(\mathbb{C}\mathbb{P}^2) \,.
\ee
The cohomology groups for the 7-dimensional space are
\be
\label{eq:cohoYpk}
H^\bullet(Y^{p,k}(\bC\P^2);\Z)=\{\Z\,, 0\,, \Z \oplus \Z_{\tn{gcd}(p,k)}\,, 0\,, \Gamma\,, \Z\,, \Z_{\tn{gcd}(p,k)}\,, \Z\}\,.
\ee
where $\Gamma$ is a finite group given by
\be\label{eq:lattice}
\Gamma \cong
\Z^2/ \langle (3k,k),(k,p)\rangle \,.
\ee
We expand $\br G_4$ on generators for each of these non-trivial elements
\be
\br G_4 = N \br \vol_{\ads{4}} +  \br F_2 \star \br v_2 + \br B_2 \star \br t_2 + \sum_{\alpha=1}^{2} \br b^{\alpha} \star \br t_4^{\alpha} \,,
\ee
where we have included the flux of the M2-branes in the first term and the parameters $\br b^\alpha$ represent the torsion $G_4$ flux.

\paragraph{$\br t_2$ generators. } We use methods described in section \ref{sec:mapbetween} to compute the 1-form symmetry generator (for cases with non-trivial $\br B_2$ field, so gcd$(p,k) \neq 1$)
\be
Z^6 = \sum_{a=1}^{p-1} a C_a \,,
\ee
where $C_a$ are compact toric divisors associated to the points $(0,0,a,1)$ in the toric diagram.

\paragraph{$\br v_2$ generators. } We have $b_2(Y^{p,k}(\bC \P^2))=1$, and we can use any one of the non-compact divisors to represent the single $\br v_2$ generator.

\paragraph{$\br t_4^{\alpha}$ generators. } We follow the prescription in section \ref{sec:mapbetween} and construct a basis of compact 4-cycles.
We wish to compute the torsional components of $\Gamma$. Again focusing on cases where $\text{gcd}(p,k) \neq 1$, we obtain the following formula from the Smith decomposition of the lattice $\langle (3k,k),(k,p) \rangle$ 
\be\label{eq:torsionalvalues}
H^4(Y^{p,k}(\mathbb{C}\mathbb{P}^2);\Z) = \Gamma =
\Z_{\text{gcd}(p,k)} \oplus \Z_{\frac{k(3p-k)}{\text{gcd}(p,k)}} \,
\ee
We generically denote these torsional components
\be
\Gamma = \Z_{k_1} \oplus \Z_{k_2} \,.
\ee
We can independently turn on $G_4$ flux of varying amounts in both directions. These flux numbers are denoted $(b^1,b^2)$, along the directions given in \eqref{eq:torsionalvalues}.

In \cite{Benini:2011cma} the authors parametrize $G_4$ flux along the $\Gamma$ directions with two integers $(n_0,n_1)$, which differ from $(b^1,b^2)$ by a basis change. To make contact with their results we require a mapping between $(n_0, n_1)$ and the torsional flux parameters introduced in the differential cohomology language above. This basis change can be read off by the column entries in the matrix $B$ defined as 
\be
\label{eq:bton}
\begin{pmatrix} \text{gcd}(p,k) & 0 \\ 0& \frac{k(3p-k)}{\text{gcd}(p,k)} \end{pmatrix} = \text{SNF} \begin{pmatrix} 3k & k \\ k & p \end{pmatrix} = A \cdot \begin{pmatrix} 3k & k \\ k & p \end{pmatrix} \cdot B \,.
\ee
We are able to provide a general expression for $Y^{p,p/c}$ for some $c \in \Z$ which divides $p$
\be
b^1 = 1\times n_0 + 0\times n_1 \,,\quad b^2 = -c\times n_0 + 1\times n_1 \,,
\ee
but give a selection of numerical values in table \ref{tab:mappings}.

\paragraph{SymTFT Coefficients.} We derive the following Symmetry TFT for general $Y^{p,k}(\mathbb{C}\mathbb{P}^2)$ geometries with $\text{gcd}(p,k)$ non-trivial \footnote{Note here we do note include $\br F_4 \star \br b \star \br b$ and $\br p_1$ terms as these contain only one or zero dynamical fields. The coefficients $\alpha$ are expressible in terms of the $\Lambda^{...}_{...}$ of \eqref{eq:CSinvariants}, but we choose this notation from now on for compactness.}
\be\label{eq:fullypksymtft}
\ba
\frac{\Stop}{2\pi}=&+ \alpha_{FF(k_1)}\int_{\cM_4} \br F_2 \star \br F_2 \star \br b^1+ \alpha_{FF(k_2)}\int_{\cM_4} \br F_2 \star \br F_2 \star \br b^2\\
&-\alpha_{FB(k_1)}\int_{\cM_4} \br F_2 \star \br b^1 \star \br B_2
-\alpha_{FB(k_2)} \int_{\cM_4} \br F_2 \star \br b^2 \star \br B_2 
\\
&- \alpha_{BB(k_1)} \int_{\cM_4} \br b^1 \star \br B_2 \star \br B_2-\alpha_{BB(k_2)} \int_{\cM_4} \br b^2 \star \br B_2 \star \br B_2 \,.
\ea 
\ee 
Given the non-trivial way the $\br t_4$ dual 4-cycles are determined, we do not expect a nice closed-form expression for general $p,k$ for all coefficients. In table \ref{tab:pkgeneralvalues2} we summarize the $\alpha$ coefficients for a large set of values of $(p,k)$. 

Since the background torsion flux participates in all the above couplings, this general SymTFT only contains terms of three types: $FF$, $BF$ and $BB$ -- we are particularly interested in the latter two\footnote{With a choice of boundary conditions the $FF$ terms are background Chern-Simons terms for 0-form global symmetry background gauge fields.}. The $BB$ term is a 1-form symmetry anomaly, whilst the $BF$ term will be crucial in understanding possible global forms of the gauge group.

\subsection{The BF-Term}
\label{sec:BFtermMtheory}
In this section we focus in particular on terms of $BF$ type, which govern the choice of gauge group in the 3d SCFT. These terms come from two sources: the first is torsion in the geometry, as introduced via non-commuting flux operators in section \ref{sec:ABJMglobalform}. The second is from background flux, both continous and discrete. The first type is standard, appearing already in $\ads{5}\times S^5$ with $N$ units of $F_5$ flux over the external space \cite{Witten:1998wy}, as well as in \eqref{eq:KS-BF} \cite{Apruzzi:2021phx}. The latter appears via terms in the differential cohomology reduction of the 11d topological terms of the form 
\be
\int_{Y_7} \br t_4 \star \br t_2 \star \br v_2 \int_{\cM_4}  \br F_2 \star \br B_2 \star \br b \,.
\ee

\paragraph{BF-terms from Non-Commuting Fluxes. }
We follow the procedure outlined in \cite{Apruzzi:2021nmk} which we applied in section \ref{sec:ABJMglobalform} to derive a new BF term:
\be
\frac{S_{\tn{BF}}}{2\pi} = \text{gcd}(p,k) \int_{\AdS_4} B_2 \wedge dB_1 \,.
\ee
The origin of the field $B_1$ is exactly the same as that presented in section \ref{sec:ABJMglobalform}, and the derivation of the coefficient follows analogously with minor modifications.

\paragraph{BF-terms from 11d CS-term. }
 
From reduction of the 11d CS-term we obtain the following 4d term of BF-type
\be
\frac{S_{\text{top}}}{2\pi} \supset \text{gcd}(p,k)\left(\alpha_{FB(k_1)} b^1 + \alpha_{FB(k_2)} b^2\right) B_2 \wedge g_2 \,.
\ee
We denote this coefficient as
\be 
\Omega^{p,k}_{n_0,n_1} \equiv \text{gcd}(p,k) \left(\alpha_{FB(k_1)}b^1 + \alpha_{FB(k_2)}b^2 \right) \,,
\ee
and give values of $\Omega^{p,k}_{n_0.n_1}$ for certain $(p,k)$ in table \ref{tab:omegacoefficients}. The coefficient depends implicitly on the integers $(n_0,n_1)$, which are linear combinations of $(b^1,b^2)$ determined by \eqref{eq:bton}. 

\paragraph{Full BF-term.}
Reduction of the 11d supergravity action on the cohomology of $Y^{p,k}(\bC\P^2)$, with discrete background flux parametrized by $(n_0,n_1)$, thus yields
\be
\label{eq:Scohoreduc}
\frac{S_{\text{BF}}}{2\pi} =  \int B_2 \wedge d \left( \text{gcd}(p,k)B_1 + \Omega^{p,k}_{n_0,n_1}c_1 \right) + \cdots \,.
\ee
Here we have left open the possibility for further BF-type terms arising when we turn on background gauge fields for the isometry group of $Y^{p,k}(\bC\P^2)$ which is $SU(3) \times U(1)^2$. In section \ref{sec:ABJM} we used a reduction to type IIA to conjecture that gauging the M-theory circle direction would furnish a new coupling with the discrete 2-form $B_2$. In appendix \ref{sec:IIAanalysis} we derive the analogous coupling from reduction to type IIA for these geometries. The result is
\be
\label{eq:fullBFTerm}\boxed{
\frac{S_{\tn{BF}}}{2\pi}= \int B_2 \wedge \left( Nf_2 + \text{gcd}(p,k)dB_1 + \Omega^{p,k}_{n_0,n_1} g_2 \right) \,,}
\ee
where $f_2=da_1$ is the field strength of the $U(1)$ 1-form gauge field associated with the M-theory circle direction. 

Let us consider the field theory interpretation of the bulk gauge fields at the level of the SymTFT -- i.e. before imposing boundary conditions consistent with the BF-coupling (\ref{eq:fullBFTerm}), which realise a particular global form of the 3d gauge group. The gauge fields in this EFT arise respectively from a reduction of $C_3$ on the free part, $c_1$ (with $g_2=dc_1$), and torsion part, $B_1$, $B_2$, of $H^\bullet(Y^{p,k};\Z)$, and from gauging the $U(1)$ isometry of the M-theory circle, $a_1$. At the boundary, the 2-form gauge field may give rise to a background for a 1-form symmetry which is a subgroup of $U(1)_{B_2}$. Fixing the 1-form gauge fields at the boundary we may realise a 0-form symmetry that sits inside $U(1)^3 = U(1)_{a_1} \times U(1)_{B_1} \times U(1)_{c_1}$. In particular, we can parametrize a set of 1-form gauge fields $A$, $A'$ for $U(1)^2 \subset U(1)^3$ defined by 
\be
\label{eq:U(1)sdecouple}
(a_1, B_1 ,c_1) = (-yA - z A',x A, x A') \,,
\ee
with 
\be\ba
N &= x \cdot \text{gcd}(N, p,k, \Omega^{p,k}_{n_0,n_1}) \,, \\
\text{gcd}(p,k) &= y \cdot \text{gcd}(N, p,k, \Omega^{p,k}_{n_0,n_1})\,, \\
\Omega^{p,k}_{n_0,n_1} &= z \cdot  \text{gcd}(N, p,k, \Omega^{p,k}_{n_0,n_1}) \,,
\ea\ee
which decouple entirely from the action (\ref{eq:fullBFTerm}) and so can always be fixed at the boundary, giving rise to a $U(1)^{(0)}\times U(1)^{(0)}$ global symmetry of the dual field theory.
The isometry group of $Y^{p,k}$ is $SU(3) \times U(1)^{2}$. In the UV, we can identify one of the $U(1)$ factors with the topological $U(1)$ symmetry of the field theory and the other with the  R-symmetry. However, as is well-known, this R-symmetry mixes with the other $U(1)$ global symmetries at the SCFT fixed point to give the superconformal R-symmetry. (The exact IR superconformal R-charge can be determined by extremization of the 3-sphere partition function \cite{Jafferis:2010un}.) The $SU(3)$ isometry group of the base $\bC\P^2$ corresponds to the baryonic $SU(3)$ that rotates the bifundamental matter in the quiver. In addition to the M-theory $U(1)$ circle direction associated with the gauge field $c_1$, we therefore have an $SU(3) \times U(1)$ isometry for which we do not turn on gauge fields. 

\subsection{Boundary Conditions and Global Symmetries}
\label{sec:bdycond}
Given the SymTFT, we can now realise all different global forms of the gauge group of the boundary theory (up to 't Hooft anomalies which obstruct certain gaugings, which we discuss shortly). Imposing a particular set of boundary conditions on the gauge fields, consistent with the action \eqref{eq:fullBFTerm}, we pick out a specific global structure for the 3d $\cN=2$ quivers.

\paragraph{Standard Boundary Conditions. }
First consider Dirichlet boundary conditions on $a_1, B_1$ and $c_1$, and Neumann on $B_2$. The action forces the boundary constraint
\be
\left(N a_1+ \text{gcd}(p,k)B_1 + \Omega^{p,k}_{n_0,n_1} c_1 \right) = 0\,.
\ee
This corresponds to a 0-form symmetry
\be
G^{(0)} \sim U(1) \times U(1) \times \Z_{\text{gcd}(N,p,k,\Omega^{p,k}_{n_0,n_1})} \,.
\ee
This global 0-form symmetry sits inside the $U(1)^3 = U(1)_{a_1} \times U(1)_{B_1} \times U(1)_{c_1}$. The two $U(1)$'s in $G^{(0)}$ can be parametrized by $A$ and $A'$ as in \eqref{eq:U(1)sdecouple}.
There is no 1-form symmetry with this choice of boundary conditions.

\paragraph{Mixed Boundary Conditions. }
Consider fixing $c_1$ and $a_1$, but letting $B_1$ be free within $\Z_n \subset \Z_{\text{gcd}(p,k)}$, with $\text{gcd}(p,k)=nn'$. This is equivalent to saying that $B_1$ is free in $\Z_{\text{gcd}(p,k)}$ modulo the relation $n' B_1=0$. The global symmetries of this choice are therefore
\be
G \sim U(1)^{(0)} \times U(1)^{(0)} \times \Z_{\text{gcd}(N,\Omega^{p,k}_{n_0,n_1},n')}^{(0)} \times \Z_{n}^{(1)}  \,.
\ee
Clearly the special case where $(n,n')=(1,\text{gcd}(p,k))$ is the `standard' choice given above. Another special case is $(n,n') = (\text{gcd}(p,k),1)$, which realizes the largest possible 1-form symmetry group. In this case, the global symmetries are
\be
G \sim U(1)^{(0)} \times U(1)^{(0)} \times \Z_{\text{gcd}(p,k)}^{(1)} \,.
\ee
Here, since $B_2,a_1$ and $c_1$ are fixed at the boundary, the following BF term
\be\label{eq:mixedanomaly}
\int B_2 \wedge d \left(N a_1 + \Omega^{p,k}_{n_0,n_1}c_1 \right) \,,
\ee
corresponds to a 3d mixed anomaly.

\paragraph{General Boundary Conditions. }
We describe a subset of the allowed boundary conditions and the resulting global symmetries in tables \ref{tab:boundaryconditions} and \ref{tab:globalsymmofboundaryconditionchoices} respectively.

\begin{table}[htbp]
\setlength{\extrarowheight}{4pt}
    \centering
    \begin{tabular}{|c|c|c|c|}
    \hline
         BC  &$a_1$ & $B_1$ & $c_1$   \\
          \hline
          \hline
          $1$&\text{D}&\text{D}&\text{N/D; Free mod} $\Z_{n'} \subset \Z_{\Omega^{p,k}_{n_0,n_1}}$\\
         \hline
         $2$& \text{D}&\text{N/D; Free mod} $\Z_{m'} \subset \Z_{\text{gcd}(p,k)}$&\text{D}\\
         \hline
          $3$&\text{N/D; Free mod} $\Z_{l'} \subset \Z_{N}$&\text{D}&\text{D} \\
         \hline
    \end{tabular}
    \caption{A selection of the possible boundary conditions consistent with the BF-action \eqref{eq:fullBFTerm}, where D: Dirichlet and N: Neumann. We take $\Omega^{p,k}_{n_0,n_1} = nn'$, $\text{gcd}(p,k) = mm'$ and $N = ll'$. `$N/D$; Free mod $\Z_q$' for a field $A$ means that the field is free to fluctuate modulo the relation $qA=0$. 
    }
    \label{tab:boundaryconditions}
\end{table}

\begin{table}[htbp]
\setlength{\extrarowheight}{4pt}
    \centering
    \begin{tabular}{|c|c|c|}
    \hline
         Boundary Condition&$G^{(0)}$& $G^{(1)}$  \\
         \hline
         \hline
         $1$ &$U(1)^2 \times \Z_{\text{gcd}(N,p,k,n')}$& $\Z_n$ \\
         \hline
         $2$ &$U(1)^2 \times \Z_{\text{gcd}(N,m',\Omega^{p,k}_{n_0,n_1})}$& $\Z_m$\\
         \hline
         $3$ &$U(1)^2 \times \Z_{\text{gcd}(l',p,k,\Omega^{p,k}_{n_0,n_1})}$& $\Z_l$\\
         \hline
    \end{tabular}
    \caption{0- and 1-form symmetries for the boundary conditions in table \ref{tab:boundaryconditions}.
    }
    \label{tab:globalsymmofboundaryconditionchoices}
\end{table}

\subsection{1-Form Symmetry Anomaly}
Using the definition of the primary invariant, the $BB$ terms of \eqref{eq:fullypksymtft} evaluate to
\be 
\left. \frac{S_{\tn{top}}}{2\pi} \right\vert_{BB}=\Omega_{BB} \int_{\cM_4} B_2 \smile B_2\,,
\ee 
with 
\be 
\Omega_{BB}=-\alpha_{BB(k_1)}b^1 - \alpha_{BB(k_2)} b^2\,.
\ee 
The coefficients $\Omega_{BB}$ depend on $p,k,b^1$ and $b^2$ (or, equivalently $p,k,n_0$ and $n_1$). This term is a 1-form symmetry anomaly: it presents an obstruction to gauging certain subgroups of the 1-form symmetry. In other words, it is an obstruction to selecting certain boundary conditions of the BF-action \eqref{eq:fullBFTerm}. Suppose gcd$(p,k)=m m'$ and we consider gauging a subgroup $\Z_{m} \subset \Z_{\tn{gcd}(p,k)}$ of the 1-form symmetry with background $B_2$. The anomaly free condition is that
\be
2 \Omega_{BB} m'^2 = 0  \mod 1 \,.
\ee
Specific coefficients of this anomaly can be computed using table \ref{tab:pkgeneralvalues2} (for a parametrization in terms of $(n_0,n_1)$ we make use of table \ref{tab:mappings} as well). 

For example, $Y^{2,2}$ with $(n_0,n_1)$ flux numbers has $\Omega_{BB}=-\frac{3}{4}n_1$. If we consider gauging the $\Z_2$ 1-form symmetry, the anomaly free condition is that $\frac{3}{2}n_1=0  \mod 1$. Hence, we can only gauge the 1-form symmetry if $n_1$ is even. In this way the torsion flux influences the possible choices of gauge group one can have for a given theory.
Furthermore, we should highlight the presence of many mixed 0-/1-form symmetry anomalies of the type demonstrated in \eqref{eq:mixedanomaly}.

\section{Comparison to the Field Theory dual to $Y^{p,k}(\mathbb{C}\mathbb{P}^2)$}
\label{sec:fieldtheorymatching}

We now compare our results with the field theory of \cite{Benini:2011cma}, which is subtle for several reasons. The proposed quiver gauge theories initially have a parity anomaly which much be quenched to ensure consistency. The authors provide several mechanisms through which this could occur. We show that ambiguity in this anomaly resolution permeates into the global symmetries of the theory: in particular the 1-form symmetry is sensitive to the anomaly cancellation mechanism one chooses.
In this section we discuss how the SymTFT can be used to constrain this problem.

\subsection{Quiver Gauge Theories}
\begin{figure}
	\begin{center} 
		\begin{tikzpicture}
		\node (g1) at (2,2) [gauge] {$U(N_1)_{k_1}$};
		\node (g2) at (4,0) [gauge] {$U(N_2)_{k_2}$};
		\node (g3) at (0,0) [gauge] {$U(N_3)_{k_3}$};
		\draw[-{Stealth}{Stealth}{Stealth}] (g1)--(g2);
		\draw[-{Stealth}{Stealth}{Stealth}] (g3)--(g1);
		\draw[-{Stealth}{Stealth}{Stealth}] (g2)--(g3);
		\end{tikzpicture}
		\caption{Quiver diagram for theory with gauge group $\Pi_{i=1}^3 U(N_i)_{k_i}$. The triple arrows denote the fact that the bi-fundamental matter fields transform in the fundamental representation of a flavor $SU(3)$.}
		\label{fig:quiver2}
	\end{center}
\end{figure}
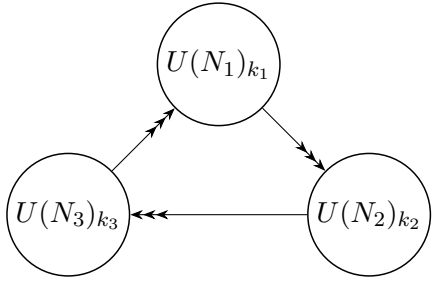

The quiver gauge theories dual to the $\AdS_4 \times Y^{p,k}(\bC\P^2)$ M-theory backgrounds \cite{Benini:2011cma} are defined for three `windows' of parameter values of the $G_4$ torsion flux, parametrized by integers $(n_0,n_1)$.
\begin{enumerate}
    \item $-k \leq n_0 \leq 0\,,\quad 0 \leq 3n_1-n_0 \leq 3p-k$
    \item $0 \leq n_0 \leq k \,, \quad 0 \leq 3n_1-n_0 \leq 3p-k$
    \item $k \leq n_0 \leq 2k \,, \quad 0 \leq 3n_1 - n_0 \leq 3p-k$
\end{enumerate}
The field theories for these three cases are 
\begin{enumerate}
    \item $U(N+n_1-p-n_0)_{-n_0+\frac{3}{2}n_1} \times U(N)_{\half n_0 - 3n_1 + \frac{3}{2}p - k} \times U(N-n_1)_{\half n_0 + \frac{3}{2}n_1 - \frac{3}{2}p+k}$
    \item $U(N+n_1-p)_{-n_0+\frac{3}{2}n_1} \times U(N)_{2n_0-3n_1+\frac{3}{2}p-k} \times U(N-n_1)_{-n_0+\frac{3}{2}n_1-\frac{3}{2}p+k}$
    \item $U(N+n_1-p)_{\half n_0 + \frac{3}{2}n_1-\frac{3}{2}q} \times U(N)_{\half n_0 - 3n_1 + \frac{3}{2}p + \half k} \times U(N-n_1 +n_0 -k)_{-n_0 + \frac{3}{2}n_1 -\frac{3}{2}p + k}$
\end{enumerate}
with bi-fundamental matter content arranged in a quiver structure shown in figure \ref{fig:quiver2}.

The theories as they are presented above suffer from a $\Z_2$ parity anomaly. The authors of \cite{Benini:2011cma} suggest that there are several mechanisms through which this residual anomaly could be cancelled. They highlight the simplest: the addition of mixed Chern-Simons couplings between the $U(1)$ pieces of different $U(N_i),U(N_j)$ factors, with levels $\Lambda_{ij}$ such that \cite{Redlich:1983dv,Redlich:1983kn,Aharony:1997bx}
\be\label{eq:anomconditions}
k_i + \half \sum_j A_{ij} N_j \in \Z \,,\quad \Lambda_{ij} - \half A_{ij} \in \Z \,.
\ee
Here $k_i$ are the Chern-Simons levels given above, $A_{ij}$ is the quiver adjacency matrix
\be
\label{eq:Amatrix}
A_{ij} = \begin{pmatrix} 0 & 3 & -3 \\ -3 & 0 & 3 \\ 3 & -3 & 0 \end{pmatrix} \,.
\ee
The first condition is satisfied by the above, but the second is not since we have so far set $\Lambda_{ij}=0$.
In \cite{Benini:2011cma}, for theories with $(n_0,n_1)=(0,0)$, the authors quote a sufficient choice
\be
\Lambda_{ij} = \begin{pmatrix}  0 & \frac{3}{2} &-\frac{3}{2} \\ \frac{3}{2} & -3 & \frac{3}{2} \\ -\frac{3}{2} & \frac{3}{2} & 0 \end{pmatrix} \,.
\ee
which does not spoil the matching of the moduli space with the geometry. Considering only the spectrum of local operators this appears to be an ambiguity in the AdS/CFT correspondence. We now return to study this ambiguity from the point of view of the 1-form symmetry, which is sensitive to $\Lambda_{ij}$. 

\subsection{1-Form Symmetry of the Quivers}
We now compute the 1-form symmetry of these field theories. The key subtlety in this computation is the presence of monopole operators which can screen Wilson lines. A monopole operator in this theory is specificed by its magnetic charges under the $U(1)$ elements of the Cartan subgroup of each $U(N_i)$ factor
\be
H_i = (m_{i,1} \dots m_{i,N_i} ) \,.
\ee 
Crucially, the choice of Chern-Simons levels $\Lambda_{ij}$ can influence the gauge charges of monopoles and therefore the 1-form symmetry. 

\paragraph{Electric Charge of Monopoles. } In a vacuum where the gauge group $\Pi_i U(N_i)$ is broken to its maximal abelian subgroup, the Lagrangian becomes \cite{Benini:2011cma}
\be\label{eq:mixedCSlagrangian}
\cL_{\text{CS}} = \sum_{i,m} \sum_{j,n} \frac{ k_i \delta_{ij} \delta_{mn} + \Lambda_{ij}}{4\pi}  A_{i,m} \wedge dA_{j,n} \,.
\ee
Suppose we put the theory \eqref{eq:mixedCSlagrangian} on $\mathbb{R} \times S^2$ and integrate over the $S^2$:
\be
\int_{S^2} \cL_{\text{CS}} = \left(\sum_{i,m} \frac{k_i}{4\pi}m_{i,m}  +\sum_{i,m} \sum_{j,n} \frac{\Lambda_{ij}}{4\pi} m_{j,n} \right)\int A_{i,m} \,.
\ee
From this we observe that a monopole acquires electric gauge charge under each $U(1)$ Cartan of each $U(N_i)$:
\be
g_{i,m} = k_i m_{i,m} + \sum_{j,l} \Lambda_{ij} m_{j,l}\,.
\ee
\paragraph{One-Loop Monopole Charge Modifications. }
This expression is modified at 1-loop \cite{Benini:2011cma} due to integrating out bifundamental matter $X_{ij}$ \footnote{Note that the formula in \cite{Benini:2011cma} contains a third correction term, which for our case of a circular quiver vanishes.}:
\be
\boxed{
g_{i,k} = k_i m_{i,k} + \delta g_{i,k} + \sum_l \sum_j \Lambda_{ij} m_{j,l} \,,}
\ee
with
\be
\delta g_{i,k} = -\half \sum_{X_{ij}} \sum_{l=1}^{N_j} \vert m_{i,k} - m_{j,l} \vert + \half \sum_{X_{ji}} \sum_{l=1}^{N_j} \vert m_{i,k} - m_{j,l} \vert \,.
\ee

\paragraph{Charge under the Center. } The charge of a monopole under the central $U(1)^3 = Z(\cG)$ is
\be
q_i = \sum_{k}^{N_i} g_{i,k} \,,
\ee
The bifundamental matter breaks this to a diagonal $U(1) \subset U(1)^3$ under which the monopole has charge
\be
q_{\text{diag}} = \sum_{i=1}^3 q_i \,.
\ee
It can be checked explicitly that one can drop the 1-loop correction ($\delta g_{i,k}$) contribution from $q_{\text{diag}}$ due to the quiver's shape:
\be
q_{\text{diag}} = \sum_{i=1}^3 q_i = \sum_{i}^3 \sum_{k=1}^{N_i} \left( k_i m_{i,k} + \sum_l \sum_j \Lambda_{ij} m_{j,l} \right)  \,.
\ee
Denoting the topological $U(1)$ charges $m_i = \sum_l m_{i,l}$, we write this as
\be
\boxed{
q_{\text{diag}} = \sum_{i=1}^3 q_i = \sum_{i}^3 \left( k_i m_i + N_i \sum_j \Lambda_{ij} m_j \right)  \,.}
\ee

\paragraph{$Y^{p,k}$ without torsion flux. } We now explicitly compute $q_{\text{diag}}$ for an arbitrary monopole in a general $Y^{p,k}$ theory without torsion flux $(n_0,n_1)=(0,0)$, with gauge group
\be
U(N-p)_{0} \times U(N)_{\frac{3}{2}p-k} \times U(N)_{-\frac{3}{2}p +k} \,.
\ee
Furthermore, we consider \textit{arbitrary} $\Lambda_{ij}$ which obeys both the parity anomaly condition and the moduli space matching condition\cite{Benini:2011cma}
\be
\Lambda_{ij} - \half A_{ij} \in \Z \,, \quad \sum_{j=1}^3 \Lambda_{ij} = 0  \,.
\ee
For a monopole with charge $(m_1,m_2,m_3)$, we obtain
\be
q_{\text{diag}} = m_1\left( -p \Lambda_{11}\right) \\
+ m_2\left(\frac{3}{2}p-k - p\Lambda_{12} \right) \\
+ m_3\left(-\frac{3}{2}p + k - p\Lambda_{13} \right) \,.
\ee
In the triangular quiver in question, the adjacency matrix is given in \eqref{eq:Amatrix}. Using the parity anomaly condition, we can rewrite ($\lambda_i \in \Z$)
\be
\Lambda_{11} = \lambda_1 \,, \quad  \Lambda_{12} = \frac{3}{2} + \lambda_2\,, \quad 
\Lambda_{13} = -\frac{3}{2} + \lambda_3 \,,
\ee
with the futher condition $\sum_i \lambda_i = 0$. We therefore have
\be
q_{\text{diag}} = -p\lambda_1 m_1 - (\lambda_2p + k)m_2 + (k+(\lambda_1+\lambda_2)p)m_3 \,.
\ee
The final 1-form symmetry of the field theory is the subgroup of the diagonal $U(1) \subset Z(\cG)$ which leaves \textit{all} monopoles invariant:
\be\boxed{
\Gamma^{(1)} = \Z_{\text{gcd}(p\lambda_1,\lambda_2p +k)} \,.}
\ee
For example, picking $\lambda_1 = 1$ and leaving $\lambda_2$ arbitrary gives $\Gamma^{(1)} =\Z_{\text{gcd}(p,k)}$. Furthermore, the choice of ${\lambda_i}$ must be compatible with supersymmetry. We have a supersymmetric solution when the effective FI parameters satisfy \cite{Benini:2011cma}
\be
\xi_1^{\text{eff}}=0 \,, \quad \xi_2^{\text{eff}} = -\xi_3^{\text{eff}} \,.
\ee
Since $\xi_1^{\text{eff}} \propto \lambda_1$, the authors of \cite{Benini:2011cma} suggest that a \textit{convenient} choice is $\lambda_1=0$, and $\lambda_2=0$. In this case, there is an enhancement of the above 1-form symmetry to $\Gamma^{(1)} = \Z_k$.
We emphasise that this solution is far from unique. Picking $\lambda_1 \neq 0$ means that we must introduce bare FI parameters $\xi_i^{\text{bare}}$ to fulfill the SUSY requirement.

\paragraph{$Y^{p,k}$ with torsion flux. }
Now consider the general $Y^{p,k}$ with torsion flux $(n_0,n_1) \neq (0,0)$. For demonstration we consider the first window of torsion space, where the gauge group is 
\be
U(N+n_1-p-n_0)_{-n_0+\frac{3}{2}n_1} \times U(N)_{\half n_0 - 3n_1 + \frac{3}{2}p - k} \times U(N-n_1)_{\half n_0 + \frac{3}{2}n_1 - \frac{3}{2}p+k} \,.
\ee
We use the parameterization (for $i \neq j$)
\be
\Lambda_{ij} = A_{ij} + \lambda_{ij}\,, \quad \lambda_{ij} \in \Z \,,
\ee
to derive
\be\ba
q_{\text{diag}}
&=m_1 \left( -(1+\Lambda_{11})n_0 + (\Lambda_{11} - \lambda_{31})n_1 - p\Lambda_{11} \right) \\
&+ m_2 \left((-1 - \lambda_{12})n_0 + (\lambda_{12} - \lambda_{32})n_1 + p(-\lambda_{12}) - k \right) \\
&+ m_3 \left( (-1 - \lambda_{13})n_0+(3 + \lambda_{13} - \Lambda_{33})n_1 + p(- \lambda_{13}) + k \right) \\
&\equiv \sum_{i=1}^3 m_i h_i \,.
\ea\ee
Once again we have the condition $\sum_j \Lambda_{ij}= 0 $ which enforces some redundancies in the parameters $\lambda_{ij}$ via $\sum_j \lambda_{ij}=0$. Since all parameters $m_i, \lambda_{ij}, \Lambda_{ii}$ are integers, the trivially acting subgroup of $U(1) \subset Z(\cG)$ is
\be\ba\label{eq:1fsgeneralypk}\boxed{
\Gamma^{(1)} = \Z_{\text{gcd}(h_1,h_2,h_3)} \,.}
\ea\ee
Again, one must check that any particular choice of $\Lambda_{ij}$ preserves supersymmetry.

\subsection{A Check on the Holographic Dictionary}
In this subsection we aim to demonstrate how consistency with the SymTFT can be used to constrain the field theory. In particular, since the 1-form symmetry is sensitive to the $U(1)$ CS-levels $\Lambda_{ij}$, the SymTFT predicts that only certain sets of $\Lambda_{ij}$ can potentially be realised. This illustrates in a concrete problem how the study of higher-form symmetries in AdS/CFT refines the dictionary. We focus on $Y^{p,p}$ with all $G_4$ torsion flux turned off. The SymTFT is 
\be\label{eq:BFexample}
\frac{S_{\text{BF}}}{2\pi} =  \int p B_2 \wedge dB_1 + NB_2 \wedge da_1  \,.
\ee
The gauge group of the 3d field theory \cite{Benini:2011cma} with which we would like to match a boundary condition of the SymTFT is 
\be
U(N-p)_{0} \times U(N)_{\half p} \times U(N)_{-\half p} \,,
\ee
with 1-form symmetry given by
\be\label{eq:fieldtheory1fsYpk}
\Gamma^{(1)} =\Z_{p \cdot \text{gcd}( \lambda_1 , \lambda_2 + 1)} \,.
\ee
We want to show that not every choice of $\lambda_1, \lambda_2$ is consistent with \eqref{eq:BFexample}. 
In more precise terms, imposing boundary conditions consistent with the BF-term can give rise to a restricted set of 1-form symmetries. We show that not all values of $\lambda_1, \lambda_2$ corresponds to field theories whose 1-form symmetry belongs to this set.

If we pick the Dirichlet boundary condition for $a_1$ and Neumann for $B_1$, the boundary field theory has 1-form symmetry $\Gamma^{(1)} = \Z_p$. Swapped boundary conditions would give $\Gamma^{(1)}=\Z_N$ 1-form symmetry, whilst any mixed condition would give $\Gamma^{(1)} \subseteq  \Z_{\tn{gcd}(p,N)} \subseteq \Z_p$. It is clearly not possible therefore to pick a boundary condition with $\Gamma^{(1)} =\Z_{l\cdot p}$, for some $l \in \mathbb{N}$ for all $N$. Noticing that the field theory result \eqref{eq:fieldtheory1fsYpk} is valid for all $N$, we can therefore constrain $\Lambda_{ij}$ to be such that
\be
l = \text{gcd}(\lambda_1, \lambda_2+1) = 1 \,.
\ee
Thus compatiblity between the SymTFT and field theory computations can be used to constrain the $U(1)$ Chern-Simons levels conjectured to resolve the known parity anomalies of these theories. We have focused here on a simple $Y^{p,p}$ model without torsion flux for concreteness, but claim that this technique is generically applicable to a broader class of examples. For general $(p,k,n_0,n_1)$ the coefficients $\Omega^{p,k}_{n_0,n_1}$ are given in table \ref{tab:omegacoefficients} and the 1-form symmetry is given by \eqref{eq:1fsgeneralypk}: with this information one can run a similar analysis in any case of interest.

\section{Outlook}
\label{sec:outlook}

There are several possible avenues of future work, some of which we summarize now:

The utility of SymTFTs is only being uncovered, and much remains to be understood, both field theoretically, but also in the realization of SymTFTs from string/M-theory. 
 In this work we derived SymTFT terms from two sources: the differential cohomology reduction of $C_3$, and from gauged isometries of the internal space $Y_7$. However, these two sources were kept distinct, whereas, ideally, they would be treated in a unified manner. If one could identify the appropriate framework, we expect that this could yield new topological couplings in many interesting setups. 

Although the examples that we considered were based on Calabi-Yau cones and associated Sasakian 7-manifolds (as well as their holographic counterparts), the methods should equally apply to other string/M-theory compactifications with special or exceptional holonomy. A natural extension of the work in this paper is to consider Spin$(7)$ holonomy spaces, which often are quotients by orientation-reversal of Calabi-Yau fourfolds. 

In view of holography, our main example was to study duals to 3d $\mathcal{N}=2$ SCFTs, and we focused on the addition of extra $U(1)$ Chern-Simons terms as a resolution to the parity anomaly of the $\cN=2$ quiver gauge theories of \cite{Benini:2011cma}. One other possibility is to restrict the gauge group to $\cG' = \left(\Pi_{i=1}^3 SU(N_i)\right) \times U(1)$. It would be interesting to examine the full scope of the SymTFT in terms of its constraining power with regards to the consistency of these field theories.

A possible future direction is to consider the SO-Sp type 3d SUSY gauge theories, which are defined as the worldvolume theory of $N$ M2-branes probing a $\mb{C}^4/\hat{D}_k$ singularity \cite{Aharony:2008ug}. Here $\hat{D}_k$ is the binary dihedral group with order $4k$, and the singularity $\mb{C}^4/\hat{D}_k$ is the anti-holomorphic involution of the toric singularity $\mb{C}^4/\mb{Z}_{2k}$. To compute the SymTFT in this case, one needs to work out a real resolution of $\mb{C}^4/\hat{D}_k$. It would be interesting to compare the geometric results with the expected higher-form, higher-group and non-invertible symmetries from field theory \cite{Beratto:2021xmn, Mekareeya:2022spm}.

The SymTFT is a powerful tool. In our analysis we have focused on two of its key features: encoding the choice of the global form of the gauge group, and the presence of 't Hooft anomalies for higher-form symmetries. However, by definition, the SymTFT in all its generality should encode all symmetry information about its associated QFT(s). Developing this further, field theoretically, and in conjunction with string/M-theory/holography provides a very exciting future research direction.

\section*{Acknowledgements}
We thank Fabio Apruzzi, Ibou Bah, Lakshya Bhardwaj, Mathew Bullimore, Cyril Closset, Federico Bonetti, Lea Bottini, I\~{n}aki Garc\'ia Etxebarria, Andrea Ferrari, Saghar Sophie Hosseini, Jingxiang Wu for discussions. 
MvB would like to acknowledge the hospitality of the Niels Bohr Institute during the later stages of this work.
SSN  acknowledges support through the Simons Foundation Collaboration on “Special Holonomy in Geometry,
Analysis, and Physics”, Award ID: 724073, Schafer-Nameki.
YNW is supported by National Science Foundation of China under Grant No. 12175004 and by Peking University under startup Grant No. 7100603667.

\appendix 

\section{Gauging Isometries}
\label{sec:isometries}
We consider the effects of gauging (a subgroup of) the isometry group $G$ of the internal space $Y_7$. The isometry 1-form gauge fields are associated with 0-form global symmetries in the 3d boundary theory, and may exhibit non-trivial topological bulk couplings with the gauge fields of the differential cohomology reduction, such as 0-/$p$-form mixed anomalies.

Suppose that $Y_7$ admits a collection of Killing vectors $k^\mu_I$ with $\mu$ a curved index and $I$ labelling a basis. There is a $G$ Lie algebra structure
\be 
\pounds_I k_J=[k_I,k_J]=\tensor{f}{_I_J^K}k_K\,,
\ee
where repeated $G$-indices are summed over. 
Gauging an internal $p$-form $x$ amounts to the substitution $x \rightarrow x^g$ where
\be 
x^g=\sum_{M=0}^p \frac{1}{M!} A^{I_1} \dots A^{I_M} \iota_{I_1} \dots \iota_{I_M} x\,,
\ee 
where $A^I$ are external connections associated to the Killing vectors $k^\mu_I$. The field strengths of the isometry gauge fields are
\be 
F^I=dA^I-\half \tensor{f}{_J_K^I}A^J \wedge A^K\,.
\ee 
Under a gauge transformation the gauge field and field strength transform as
\be 
\label{eq:gaugetransA}
\delta_\lambda A^I=-d\lambda^I-\tensor{f}{_J_K^I}\lambda^J A^K\,, \qquad \delta_\lambda F^I=-\tensor{f}{_J_K^I}\lambda^J F^K\,.
\ee 
A gauged internal form $x^g$ in an arbitrary representation of the isometry algebra transforms as
\be 
\delta_\lambda(x^g)=\lambda^J(\pounds_J x)^g\,.
\ee

\subsection{Equivariant Cohomology}
The approach we will now take is to uplift the flux to $G$-equivariant cohomology $H^p_G(\cM_{11};\R)$, see e.g. \cite{Libine} for an introduction. This description is at the level of differential forms and so is only suited for describing the free part of the cohomology. The method was used in \cite{Bah:2019rgq} to compute anomaly polynomials for even dimensional QFTs that are engineered from 11d supergravity with 4-form background flux over internal 4-cocycles, i.e. 
with $\cL=0$ and $\cN^i \neq 0$ in \eqref{eq:G4bg}. They also considered gauging of internal harmonic 2-forms in $G_4$. The key idea is that under gauging the flux picks up a dependence on the external connections $A^I$ so that $G_4^g$ is generically neither closed nor gauge invariant. In \cite{Bah:2019rgq} it was shown that such a 4-form flux can be given a completion $G_4^g \rightarrow G_4^\tn{eq}$ that allows for a lift to $G$-equivariant cohomology, i.e. such that
\be 
\delta_\lambda G_4^\tn{eq}=0\,, \qquad d G_4^\tn{eq}=0\,,
\ee 
and that this completion correctly computes the anomaly polynomial for the isometry gauge fields.

We are mainly interested in backgrounds associated with the presence of M2-branes, which are supported by flux over the external space $\cM_4$. 
For now we therefore take a general perspective, allowing for any background flux configuration. We will derive the equivariant completion of $G_4$ expanded on all harmonic $p$-forms of the internal geometry $Y_7$
and compute the resulting contribution to the SymTFT.

The free part of the background flux $\cL \vol_{\cM_4}$, is clearly invariant under gauging of the isometries of the internal space, so any $A^I$ dependence in the free part of the flux is necessarily associated with the expansion on the cohomology of $Y_7$. That is, the harmonic $p$-forms $\omega_p^i$, $i=1,\dots,b^p(Y_7)$ may transform under gauging $\omega_p^i \rightarrow (\omega_p^i)^g$. (In the following we will suppress indices $i,j=0,\dots, b^p(Y_7)$ that label the basis of the $p$'th cohomology in an attempt to keep notation to a minimum).
The equivariant completion of the 4-form flux can therefore be parametrized as
\be 
\label{eq:equiexp}
G_4^\tn{eq}=\cL \vol_{\cM_4}+g_4+g_3 \wedge  \omega_1^\tn{eq}+g_2 \wedge \omega_2^\tn{eq}+g_1 \wedge \omega_3^\tn{eq}+\cN \omega_4^\tn{eq}\,,
\ee 
where an ansatz for the putative $p$-form representatives of the equivariant cohomology
$H_G^p (Y_7;\Z)$ is
\be\ba
\omega_1^{\text{eq}} &= \omega_1^g \,, \\
\omega_2^\text{eq} &=\omega_2^g+ F^I \eta_I \,, \\
\omega_3^{\text{eq}} &= \omega_3^g + F^I \wedge \eta_{1I}^g \,, \\
\omega_4^{\text{eq}} &= \omega_4^g + F^I \wedge \eta_{2I}^g  + \epsilon_{IJ} F^I \wedge F^J  \,,
\ea\ee
with $\eta_{2I}$, $\eta_{1I}$, $\eta_I$, $\varepsilon_{IJ}=\varepsilon_{JI}$ internal 2-, 1-, and 0-forms with isometry group indices, which do not have to be closed. 

The terms linear in the $F^I$ transform as
\be 
\delta_\lambda(F^I \wedge \eta_{pI}^g)=-\tensor{f}{_J_K^I}\lambda^J F^K \wedge \eta_{pI}^g+F^I \wedge \lambda^J(\pounds_J \eta_{pI})^g\,,
\ee 
and the quadratic piece gives
\be 
\delta_\lambda(\epsilon_{IJ}F^I \wedge F^J)=-\epsilon_{IM} \tensor{f}{_J_K^I}\lambda^JF^K \wedge F^M- \epsilon_{IM}F^I \wedge \tensor{f}{_J_K^M}\lambda^J F^K+\lambda^J (\pounds_J\epsilon_{IM}) F^I \wedge F^M  \,,
\ee 
Requiring gauge invariance of $G_4^\tn{eq}$ thus implies
\be  
\label{eq:gaugetransw}
\pounds_I \eta_{pJ}=\tensor{f}{_I_J^K} \eta_{pK}\,, \qquad \pounds_I \epsilon_{JK}=\tensor{f}{_I_J^M} \epsilon_{MK}+\tensor{f}{_I_K^M} \epsilon_{JM}\,.
\ee 
To obtain the conditions for $G_4^\tn{eq}$ to be closed, we will make use of the identity \cite{Bah:2019rgq}
\be 
\label{eq:idg}
d(x^g)+A^I \wedge (\pounds_I x)^g=(dx)^g+(\iota_I x)^g \wedge F^I\,,
\ee 
which implies
\be 
d(\omega_p^g)=(\iota_I \omega_p)^g \wedge F^I\,,
\ee 
for the harmonic forms.
Furthermore, we have
\be 
d(F^I \wedge \eta_{pI}^g)= F^I \wedge \lbb (d\eta_{pI})^g+(\iota_J \eta_{pI})^g \wedge F^J -A^J \wedge (\tensor{f}{_J_I^K} \eta_{pK})^g\rbb+dF^I \wedge \eta_{pI}^g\,.
\ee 
Observe that
\be 
dF^I= \tensor{f}{_J_K^I}A^J \wedge F^K\,,
\ee 
where we used the Jacobi identity. Thus, we find
\be 
d(F^I \wedge \eta_{pI}^g)= F^I \wedge (d\eta_{pI})^g+(\iota_I \eta_{pJ})^g \wedge F^I \wedge F^J\,.
\ee 
For the quadratic term we have
\be
d(\epsilon_{IJ}^g F^I \wedge F^J)= (d \epsilon_{IJ})^g \wedge F^I \wedge F^J\,,
\ee 
since all other contributions vanish on dimensional grounds. Hence, $G_4^\tn{eq}$ is closed if $\eta_{pI}$ and $\epsilon_{IJ}$ solve
\be \label{eq:internalformconstraints}
\ba 
\iota_I \omega_4+ d \eta_{2I}&=0\,, \qquad &&\iota_{(I} \eta_{2J)}+ d\epsilon_{IJ}=0\,,\\
\iota_I\omega_3+d \eta_{1I}&=0\,, \qquad
 &&\iota_I\omega_2+ d \eta_I=0\,.
\ea 
\ee 
Notice that it is {\it not} always possible to solve this set of equations. Equivariantising a harmonic $p$-form is obstructed if $\iota_I \omega_p$ is not exact\footnote{We thank Ibou Bah and Federico Bonetti for pointing out this subtlety.}. This issue is circumvented e.g. if the even/odd cohomology of $Y_7$ is trivial, as is the case in all the examples we consider in this paper.

We would like to interpret the equivariant completions $\omega_p^\tn{eq}$ as equivariant maps 
\be
f: \mathfrak{g} \to \Omega^\bullet(Y_7) \,,
\ee
which obey
\be 
\label{eq:equitrans}
\pounds_I f(\cX) =  \tensor{f}{_I_J^K}\cX^J \frac{\partial}{\partial \cX^K} f(\cX) \,,
\ee
and are closed under 
\be 
\label{eq:equiclose}
d_\tn{eq}(F(\cX)) \equiv d(f(\cX)) + \iota_{\cX} f(\cX)\,,
\ee 
for $X = \cX^I t_I \in \mathfrak{g}$ in some basis $\{t_I\}$ of $\mathfrak{g}$.
Given this identification, the forms $\omega_p^\tn{eq}$ correspond to non-trivial elements in $G$-equivariant cohomology. Since the expressions derived for $\omega_4^{\text{eq}}$ and $\omega_2^{\text{eq}}$ match those in \cite{Bah:2019rgq}, we need only consider the odd pieces $\omega_3^{\text{eq}}$ and $\omega_1$. We consider the map
\be
f_{\omega_3}: \cX^I \to f_{\omega_3}(\cX^I) = \omega_3 + \cX^I \eta_{1I} \,,
\ee
for some $X = \cX^I t_I \in \mathfrak{g}$. 
We begin with closure, and in particular identify $F^I = \cX^I$ 
\be\ba
d_{\text{eq}} f_{\omega_3} (F) &=  d(f(F)) + F^I \wedge \iota_I f(F) \\
&= d\left(\omega_3 + F^I \wedge \eta_{1I} \right) + F^I \wedge \iota_I \left(\omega_3 + F^J \wedge \eta_{1J} \right) \,, \\
&= d\omega_3 +dF^I \wedge \eta_{1I} + F^I \wedge d\eta_{1I} + F^I \wedge \iota_I \omega_3 +F^I \wedge F^J \wedge \iota_I \eta_{1J} \,.
\ea\ee
which reproduces the condition in \eqref{eq:internalformconstraints}. Furthermore, it implies a new condition 
\be 
\iota_{(I} \eta_{1J)}=0\,,
\ee 
which is not necessary for $g_1 \wedge \omega_3^\tn{eq}$ to be closed, but it {\it is} necessary to be able to identify $\omega_3^\tn{eq}$ with a non-trivial $G$-equivariant cohomology class.
The equivariant transformation \eqref{eq:equitrans} reproduces the result for the Lie derivative of $\eta_{1I}$ in \eqref{eq:gaugetransw}. The map we consider for the harmonic 1-form is simply
\be
f_{\omega_1}\,: \cX^I \rightarrow f_{\omega_1}(\cX^I)=\omega^1\,,
\ee 
which trivially satisfies \eqref{eq:equitrans}. However, from \eqref{eq:equiclose} we find
\be 
\iota_I \omega_1=0\,,
\ee 
which is not required for $G_4^\tn{eq}$ to be closed, but without it $\omega_1$ cannot represent an element of $H^1_G(Y_7;\R)$. Note that this is a constraint on the geometry, similar to the condition discussed after \eqref{eq:internalformconstraints}. If this condition is not satisfied it poses an obstruction to an equivariant uplift.

Finally, before computing the SymTFT, we make a few comments on the scope of this formalism and the interplay with torsion. The limitation of the description of equivariant cohomology in terms of differential forms is of course that it only captures the free part of the cohomology of $\cM_{11}$, since the torsion generators are not associated with any differential form. In principle, a complete treatment of the gauging of isometries in the presence of torsion might be formulated in {\it equivariant differential cohomology} $\hat{\mathbb{H}}^n_G(M;\Z)$. However at present, it not known how to write down an appropriate parametrization of $\hat G_4 \in \hat{\mathbb{H}}^4_G(M;\Z)$, and how to define an action principle generalising \eqref{eq:defStop}.
So, whereas the differential cohomology formulation of section \ref{sec:SymTFT} determines the couplings of the KK field strengths $g_p^i$ and discrete gauge fields $B_p^\alpha$, 
and the equivariant cohomology uplift discussed above gives the couplings of the $g_p^i$ and $F^I$, we are not sensitive to couplings involving both $B_p^\alpha$ and $F^I$ fields. Yet, in section \ref{sec:ABJM} we find evidence from the AdS/CFT correspondence for the existence of such couplings.

\subsection{SymTFT with Gauged Isometries}
\label{sec:gaugedisofree}

We now compute the SymTFT couplings of the isometry gauge fields by dimensionally reducing the 12-dimensional M-theory gauge invariant polynomial
\be
I_{12} = - \frac{1}{6} (G_4^\tn{eq})^3 - G_4^\tn{eq} \wedge X_8 \,.
\ee
Using \eqref{eq:equiexp} and restoring indices so that
\be 
G^\text{eq}_4= \sum_{p=0}^4 \sum_{i=1}^{b^p(Y_7)}g_{4-p}^i \wedge (\omega_{p}^i)^\text{eq}\,,
\ee 
the CS-term gives
\be\ba
-\frac{1}{6}\int_{Y_7} (G_4^{\text{eq}})^3-G_4^3= 
\sum_{ijk} & \lb -\half H_{I}^{ijk} g_1^i \wedge g_1^j \wedge g_1^k +\Theta_I^{ijk} \cN^i g_1^j \wedge g_2^k + \mathcal{H}_{I}^{ijk} \cN^i \cN^j g_3^k \rb \wedge F^I\\
+\sum_{ijk} & \Pi_{IJ}^{ijk} \cN^i \cN^j  g_1^k \wedge F^I \wedge F^J\,.
\ea\ee
The new coefficients are given by
\be 
\ba 
H_{I}^{ijk}&= \int_{Y_7} \eta_{1I}^i \wedge \omega_3^j \wedge \omega_3^k\,, \qquad && \Theta_I^{ijk}=\int_{Y_7}  \omega_4^i \wedge \omega_2^k \wedge \eta_{1J}^j+\eta_I^k \omega_4^i \wedge \omega_3^j \,, \\
\mathcal{H}_{I}^{ijk} &= \int_{Y_7} \omega_4^i \wedge \eta_{2I}^j \wedge \omega_1^k \,, \qquad && \Pi_{IJ}^{ijk}=\int_{Y_7}  \half \eta_{2I}^i \wedge \eta_{2J}^j \wedge \omega_3^k+\omega_4^i \wedge \eta_{2I}^j \wedge \eta_{1J}^k+\epsilon_{IJ}^j \omega_4^i \wedge \omega_3^k\,.
\ea \ee
Furthermore, equivariantly completing $X_8$ amounts to
\be
p_1(TY_7) \to p_1(TY_7)^g + F^I \sigma_{2I}^g + F^I F^J \phi_{IJ} \,.
\ee
These new terms will be subject to constraint equations.
However, we previously argued that only $p_1(TY_7) \wedge p_1(T\cM_4)$ survives inside $X_8$, then on dimensional grounds we do not obtain any new terms from the isometries.

Finally, we can also have topological contributions from the kinetic term. We find that the kinetic term contains new contributions to the SymTFT
\be
\label{eq:isokin}
\int_{Y_7} \half \vert G_4^\tn{eq} \vert ^2 \subseteq \cL g_4+\lbb \int_{Y_7} \eta_{I} \vol_{Y_7} \rbb \cL g_2 	\wedge F^I + \sum_i \lbb \int_{Y_7} \epsilon_{IJ}^i \vol_{Y_7} \rbb \cL \cN^i F^I \wedge F^J   \,.
\ee
Here we have used a normalisation where $Y_7$ has unit volume.

\section{BF-Terms from Type IIA for $Y^{p,k}$}
\label{sec:IIAanalysis}
In this section we utilize a reduction to  type IIA to derive an extra BF-term contribution on top of those computed via M-theory methods in section \ref{sec:BFtermMtheory}. In \cite{Benini:2011cma} the authors reduce the M-theory solution on AdS$_4 \times Y^{p,k}$ background to IIA along a circle. The IIA supergravity background is
\be
\text{AdS}_4 \times_w M_6 \,,
\ee
where $M_6$ is a $S^2$ bundle over $\mathbb{C}\mathbb{P}^2$. The homology groups of $M_6$ are 
\be
H_{\bullet} = \{\Z,0,\Z^2,0,\Z^2,0,\Z\} \,.
\ee
The RR field strengths and Kalb-Ramond field are parametrized as
\be\ba
\relax [F_2] &= pD^+ -kD \,, \\
[B_{\tn{NS}}] &= - b_0D^- +b^+D\,,\\
[F_6] &= N D \cdot \cC^+ \,.
\ea\ee
Here, $\{D,D^+,D^-\}$ are an over-complete basis of 4-cycles. There is a dual set of 2-cycles $\{ \cC, \cC^+, \cC^- \}$ which is also overcomplete. They are related by
\be
D^+ = D^- + 3D \,, \quad \cC^+ = \cC^- + 3\cC \,.
\ee
Their mutual intersections are given in table \ref{tab:intersectionsofbasisforIIAgeometry}.
\begin{table}[]
    \centering
    \begin{tabular}{|c|c c c| c c c|}
    \hline
         &$\cC$&$\cC^+$&$\cC^-$&$D$&$D^+$&$D^-$  \\
         \hline
         \hline
         $D$&$0$&$1$&$1$&$\cC$&$\cC^+$&$\cC^-$  \\
         $D^+$&$1$&$3$&$0$&$\cC^+$&$3\cC^+$&$0$  \\
         $D^-$&$1$&$0$&$-3$&$\cC^-$&$0$&$-3\cC^-$  \\
         \hline
    \end{tabular}
    \caption{Intersections between 4-cycles $\{D,D^+,D^-\}$ and 2-cycles $\{\cC,\cC^+,\cC^-\}$ \cite{Benini:2011cma}. }
    \label{tab:intersectionsofbasisforIIAgeometry}
\end{table}
We write a set of Poincar\'e dual 2- and 4-forms
\be\ba
\{D,D^+,D^- \} &\leftrightarrow \{\omega_2,\omega_2^+,\omega_2^-\} \,,\\
\{\cC,\cC^+,\cC^- \} &\leftrightarrow \{\omega_4,\omega_4^+,\omega_4^-\} \,.
\ea\ee
Let us consider fluctuations around this background in this basis
\be\ba
F_2'&=F_2 + f_2 = p\omega_2^+ -k\omega_2 + f_2  \,, \\
B_{\tn{NS}}'&=B_{\tn{NS}} + b_2 = - b_0\omega_2^- +b^+\omega_2 + b_2 \\
F_6'&=F_6 +f_6 = N \omega_2 \wedge \omega_4^+ + g_4^+\wedge \omega_2^+ +  g_4^{(0)} \wedge \omega_2 + g_2^- \wedge \omega_4^- + g_2^{(0)} \wedge \omega_4 \,.
\ea\ee
We now look for single derivative terms in the IIA equations of motion \cite{Apruzzi:2021phx,Apruzzi:2021nmk}, which will dominate at long distances, i.e. near the conformal boundary of AdS. In particular, we are interested in couplings involving $b_2$ and 1-form gauge fields
\be\ba\label{eq:IIAeom}
d \star_{10} F_2 &= H_3 \wedge F_6 = db_2 \wedge N \text{vol}(M_6) + \cdots = Ndb_2 \wedge \text{vol}(M_6) + \cdots \,. \\
d \star_{10} H_3 &= F_2 \wedge F_6 = \left( pg_2^{(0)} - kg_2^-  + Nf_2 \right) \wedge \text{vol}(M_6) + \cdots  \\
d\star_{10} F_6 &= H_3 \wedge F_2 = db_2 \wedge (p\omega_2^+ - k\omega_2) + \cdots \\
\ea\ee
The Bianchi identities are $dF_6 = H_3 \wedge F_4$, $dF_2 = H_3 \wedge F_0$, $dH_3=0$ are trivially satisfied given our expansion. At the boundary, we are left with the following topological equations of motion
\be\ba
Ndb_2 &=0 \\
pdb_2 &= 0 \\
kdb_2 &= 0 \\
\left( pg_2^{(0)} - kg_2^- + Nf_2 \right) &= 0 \,.
\ea\ee
These equations of motion are reproduced by 
\be
\frac{S_{\tn{IIA}}}{2\pi}=  \int b_2 \wedge \left( Nf_2 + pg_2^{(0)} -kg_2^- \right) \,.
\ee
If we package up
\be\label{eq:packaging}
(3p-k)g_2^+ + pg_2^{(0)} \equiv \text{gcd}(p,k) \left( q_1 g_2^- + q_2 g_2^{(0)} \right) \equiv \text{gcd}(p,k) \widetilde{g}_2 \,,
\ee
we can rewrite 
\be
\frac{S_{\tn{IIA}}}{2\pi} = \int b_2 \wedge \left( Nf_2 + \text{gcd}(p,k) \widetilde{g}_2 \right) \,.
\ee
Let us compare with the M-theory analysis of section \ref{sec:BFtermMtheory}, in particular \eqref{eq:Scohoreduc}. 
As discussed in section \ref{sec:ABJMglobalform}, we conjecture that $b_2$, which couples electrically to the fundamental string, uplifts to $B_2$ which couples to M2-branes wrapping the torsional 1-cycle. The field $a_1$ sourced by D0-branes with field strength $f_2$ uplifts to the $U(1)$ isometry gauge field $A_1$ associated with the M-theory circle direction. In IIA, the 1-form gauge field $\widetilde{c_1}$ with field strength $\widetilde{g_2}$ couples electrically to D4-branes wrapping the two 4-cycles in the $M_6$ geometry. We expect that the linear combination $\widetilde{c_1}$ maps to $B_1$ upon uplift to M-theory, which couples electrically to M5-branes wrapping the torsional 5-cycle.

The $N B_2 \wedge f_2$ coupling is precisely the one we do not have access to from M-theory. We claim that this would be visible if we combined the equivariant cohomology description with differential cohomology, analogously to the matching we did in the ABJM example. On the other hand, IIA does not see the $\Omega^{p,k}_{n_0,n_1} B_2 \wedge g_2$ term of \eqref{eq:Scohoreduc}, at the level of our analysis. We use this to conjecture an additional term in the M-theory BF-coupling:
\be
\frac{S_{\tn{BF}}}{2\pi}= \int B_2 \wedge \left( Nf_2 + \text{gcd}(p,k)dB_1 + \Omega^{p,k}_{n_0,n_1} g_2 \right) \,.
\ee

\begin{table}[H]
\setlength{\extrarowheight}{15pt}
    \centering
    \begin{tabular}{|c|c||c|c|c|c|c|c|}
    \hline
         $p$&$k$&$\alpha_{BB(k_1)}$&$\alpha_{BB(k_2)}$&$\alpha_{FB(k_1)}$&$\alpha_{FB(k_2)}$\\
         \hline
         \hline
 2&2&$\frac{3}{4}$ &$ \frac{3}{4}$ &$ \frac{1}{2} $&$ 0$ \\
 \hline
3&3&$ \frac{5}{6}$ & $\frac{2}{3}$ &$ \frac{1}{3}$ &$ 0$  \\
\hline
 4&2&$0$ &$ \frac{1}{4}$ &$ 0 $&$ \frac{1}{2} $\\
\hline
4&4&$ \frac{7}{8}$ &$ \frac{5}{8}$ &$ \frac{1}{4}$ &$ 0$ \\
\hline
4&6&$0$&$\frac{1}{4}$&$0$&$\half$ \\
\hline
5&5&$ \frac{9}{10} $&$ \frac{3}{5}$ &$ \frac{1}{5} $&$ 0$   \\
\hline
6&2& $\frac{3}{4}$ & $\frac{1}{2}$ & $\frac{1}{2}$ & $\frac{1}{2}$   \\
\hline
 6&3&$\frac{2}{3}$ & $\frac{5}{6}$ &$ \frac{2}{3} $&$ \frac{2}{3} $
   \\
   \hline
6&4&$ \frac{3}{4} $&$ \frac{1}{4}$ & $\frac{1}{2} $&$ 0$  \\
\hline
 6&6&$\frac{11}{12}$ & $\frac{7}{12} $&$ \frac{1}{6} $& $0$  \\
 \hline
6&8& $\frac{3}{4}$ &$ \half$ &$ \half $&$ \half $ \\
\hline
6&9&$0$&$\frac{1}{6}$&$\frac{2}{3}$&$\frac{2}{3}$\\
\hline
7&7& $\frac{13}{14}$ & $\frac{4}{7}$&$ \frac{1}{7}$ & $0$  \\
\hline
8&2& $0$ & $\frac{3}{4}$ & $0$ &$ \frac{1}{2}$ \\
\hline
8&4& $0$ & $\frac{5}{8}$ &$ \frac{1}{2}$ &$ \frac{3}{4}$ \\
\hline
8&6& $0$ & $\frac{3}{4}$ & $0$ & $\frac{1}{2}$ 
   \\
   \hline
8&8& $\frac{15}{16}$ & $\frac{9}{16}$ & $\frac{1}{8}$&$0$   \\
\hline
8&10& $0$ & $\frac{3}{4}$ & $0$&$\half$   \\
\hline
8&12& $0$ & $\frac{1}{8}$ & $\frac{1}{2}$ & $\frac{3}{4}$   \\
\hline
    \end{tabular}
    \caption{A selection of the $Y^{p,k}(\mathbb{C}\mathbb{P}^2)$ SymTFT coefficients obtained for selected $p,k$ values with non-trivial gcd$(p,k)$. Note we have also not included pairs of $\text{gcd}(p,k)$ values related by $(p,k) \to (p,3p-k)$.}
    \label{tab:pkgeneralvalues2}
\end{table}

\begin{table}[H]
\setlength{\extrarowheight}{10pt}
    \centering
    \resizebox{\textwidth}{!}{
    \begin{tabular}{|c||c|c|c|c|c|c|c|c|}
    \hline
         $p$&6&6&6&8&8&8&9&9 \\
         \hline
         $k$&4&8&9&6&10&12&6&12\\
         \hline
         \hline
         $b^1$&$n_1-n_0$&$3n_1-2n_0$&$2n_1-n_0$&$n_1-n_0$&$4n_1-3n_0$&$2n_1-n_0$&$n_1-n_0$&$3n_1-2n_0$\\
         \hline
         $b^2$&$2n_1-3n_0$&$4n_1-3n_0$&$3n_1-2n_0$&$3n_1-4n_0$&$5n_1-4n_0$&$3n_1-2n_0$&$2n_1-3n_0$&$4n_1-3n_0$\\
         \hline
    \end{tabular}}
    \caption{Mappings from $(n_0,n_1)$ torsion flux numbers to $(b^1,b^2)$ flux numbers. Here we give cases where $k \neq \frac{p}{c}$ for some $c \in \Z$.}
    \label{tab:mappings}
\end{table}

\begin{table}[H]
\setlength{\extrarowheight}{5pt}
    \centering
    \resizebox{\textwidth}{!}{
    \begin{tabular}{|c||c|c|c|c|c|c|c|}
    \hline
         $p$&$p$&$4$&4&6&6&6&6\\
         \hline
 $k$&$p$ &2&6&2&3&4&8  \\
 \hline
 \hline
 $\Omega^{p,k}_{n_0,n_1}$&$n_0$&$-2n_0+n_1$&$-2n_0+3n_1$&$-2n_0+n_1$&$-2n_0+2n_1$&$n_1-n_0$&$7n_1-5n_0$\\
\hline
\hline
$p$&6&8&8&8&8&8& \\
\hline
$k$&9&2&4&6&10&12& \\
\hline
\hline
$\Omega^{p,k}_{n_0,n_1}$ & $10n_1-6n_0$&$-4n_0+n_1$&$-4n_0+n_1$&$3n_1-4n_0$&$5n_1-4n_0$&$13n_1-8n_0$ &\\
\hline
    \end{tabular}}
    \caption{Values of $\Omega^{p,k}_{n_0,n_1}$ coefficients for various values of $p$ and $k$. These were computed by using values in table \ref{tab:pkgeneralvalues2} and mapping $(b^1,b^2)$ to $(n_0,n_1)$ (table \ref{tab:mappings}). Notice in the first column we give a general expression for $Y^{p,p}$.}
    \label{tab:omegacoefficients}
\end{table}

\bibliographystyle{JHEP}

\bibliography{BibHFS}

\end{document}